
\newbox\SlashedBox 
\def\slashed#1{\setbox\SlashedBox=\hbox{#1}
\hbox to 0pt{\hbox to 1\wd\SlashedBox{\hfil/\hfil}\hss}{#1}}
\def\hboxtosizeof#1#2{\setbox\SlashedBox=\hbox{#1}
\hbox to 1\wd\SlashedBox{#2}}

\def\mathslashed#1{\setbox\SlashedBox=\hbox{$#1$}
\hbox to 0pt{\hbox to 1\wd\SlashedBox{\hfil/\hfil}\hss}#1}

\def\ifsmall{\iffalse}  
\def\titlepagefont{}  

\def\DefineTeXgraphics{%
\special{ps::[global] /TeXgraphics { } def}}  

\def\today{\ifcase\month\or January\or February\or March\or April\or May
\or June\or July\or August\or September\or October\or November\or
December\fi\space\number\day, \number\year}
\def\eatPrefix19{}
\def\Year{\expandafter\eatPrefix\the\year}
\newcount\hours \newcount\minutes
\def\monthname{\ifcase\month\or
January\or February\or March\or April\or May\or June\or July\or
August\or September\or October\or November\or December\fi}
\def\shortmonthname{\ifcase\month\or
Jan\or Feb\or Mar\or Apr\or May\or Jun\or Jul\or
Aug\or Sep\or Oct\or Nov\or Dec\fi}

\def\TimeStamp{\hours\the\time\divide\hours by60%
\minutes -\the\time\divide\minutes by60\multiply\minutes by60%
\advance\minutes by\the\time%
${\rm \shortmonthname}\cdot\if\day<10{}0\fi\the\day\cdot\the\year%
\qquad\the\hours:\if\minutes<10{}0\fi\the\minutes$}




\def\Title#1{%
\vskip 1in{\titlefont\centerline{#1}}\vskip .5in}
 
\def\Date#1{\leftline{#1}\tenrm\supereject%
\global\hsize=\hsbody\global\hoffset=\hbodyoffset%
\footline={\hss\tenrm\folio\hss}}

\newif\ifdraftmode
\newif\ifleftlabels  

\def\nolabels{\def\wrlabeL##1{}\def\eqlabeL##1{}\def\reflabeL##1{}}
\def\writelabels{\def\wrlabeL##1{\leavevmode\vadjust{\rlap{\smash%
{\line{{\escapechar=` \hfill\rlap{\sevenrm\hskip.03in\string##1}}}}}}}%
\def\eqlabeL##1{{\escapechar-1\rlap{\sevenrm\hskip.05in\string##1}}}%
\def\reflabeL##1{\noexpand\rlap{\noexpand\sevenrm[\string##1]}}}
\def\writeleftlabels{\def\wrlabeL##1{\leavevmode\vadjust{\rlap{\smash%
{\line{{\escapechar=` \hfill\rlap{\sevenrm\hskip.03in\string##1}}}}}}}%
\def\eqlabeL##1{{\escapechar-1%
\rlap{\sixrm\hskip.05in\string##1}%
\llap{\sevenrm\string##1\hskip.03in\hbox to \hsize{}}}}%
\def\reflabeL##1{\noexpand\rlap{\noexpand\sevenrm[\string##1]}}}
\nolabels

\newdimen\fullhsize
\newdimen\hstitle
\hstitle=\hsize 
\newdimen\hsbody
\hsbody=\hsize 
\newdimen\hbodyoffset
\hbodyoffset=\hoffset 
\newbox\leftpage
\def\abstract#1{#1}
\def\rotated{\special{ps: landscape}
\magnification=1000  
\baselineskip=14pt
\global\hstitle=9truein\global\hsbody=4.75truein
\global\vsize=7truein\global\voffset=-.31truein
\global\hoffset=-0.54in\global\hbodyoffset=-.54truein
\global\fullhsize=10truein
\def\DefineTeXgraphics{%
\special{ps::[global] 
/TeXgraphics {currentpoint translate 0.7 0.7 scale
              -80 0.72 mul -1000 0.72 mul translate} def}}
\let\lr=L
\def\ifsmall{\iftrue}
\def\titlepagefont{\twelvepoint}
\trueseventeenpoint
\def\almostshipout##1{\if L\lr \count1=1
      \global\setbox\leftpage=##1 \global\let\lr=R
   \else \count1=2
      \shipout\vbox{\hbox to\fullhsize{\box\leftpage\hfil##1}}
      \global\let\lr=L\fi}

\output={\ifnum\count0=1 
 \shipout\vbox{\hbox to \fullhsize{\hfill\pagebody\hfill}}\advancepageno
 \else
 \almostshipout{\leftline{\vbox{\pagebody\makefootline}}}\advancepageno 
 \fi}

\def\abstract##1{{\leftskip=1.5in\rightskip=1.5in ##1\par}} }

\def\linemessage#1{\immediate\write16{#1}}

\global\newcount\secno \global\secno=0
\global\newcount\appno \global\appno=0
\global\newcount\meqno \global\meqno=1
\global\newcount\subsecno \global\subsecno=0
\global\newcount\figno \global\figno=0

\newif\ifAnyCounterChanged
\let\terminator=\relax
\def\normalize#1{\ifx#1\terminator\let\next=\relax\else%
\if#1i\aftergroup i\else\if#1v\aftergroup v\else\if#1x\aftergroup x%
\else\if#1l\aftergroup l\else\if#1c\aftergroup c\else%
\if#1m\aftergroup m\else%
\if#1I\aftergroup I\else\if#1V\aftergroup V\else\if#1X\aftergroup X%
\else\if#1L\aftergroup L\else\if#1C\aftergroup C\else%
\if#1M\aftergroup M\else\aftergroup#1\fi\fi\fi\fi\fi\fi\fi\fi\fi\fi\fi\fi%
\let\next=\normalize\fi%
\next}
\def\makeNormal#1#2{\def\doNormalDef{\edef#1}\begingroup%
\aftergroup\doNormalDef\aftergroup{\normalize#2\terminator\aftergroup}%
\endgroup}

\def\warnIfChanged#1#2{%
\ifundef#1
\else\begingroup%
\edef\oldDefinitionOfCounter{#1}\edef\newDefinitionOfCounter{#2}%
\ifx\oldDefinitionOfCounter\newDefinitionOfCounter%
\else%
\linemessage{Warning: definition of \noexpand#1 has changed.}%
\global\AnyCounterChangedtrue\fi\endgroup\fi}

\def\Section#1{\global\advance\secno by1\relax\global\meqno=1%
\global\subsecno=0%
\bigbreak\bigskip
\centerline{\twelvepoint \bf %
\the\secno. #1}%
\par\nobreak\medskip\nobreak}
\def\tagsection#1{%
\warnIfChanged#1{\the\secno}%
\xdef#1{\the\secno}%
\ifWritingAuxFile\immediate\write\auxfile{\noexpand\xdef\noexpand#1{#1}}\fi%
}
\def\section{\Section}
\def\Subsection#1{\global\advance\subsecno by1\relax\medskip %
\leftline{\bf\the\secno.\the\subsecno\ #1}%
\par\nobreak\smallskip\nobreak}
\def\tagsubsection#1{%
\warnIfChanged#1{\the\secno.\the\subsecno}%
\xdef#1{\the\secno.\the\subsecno}%
\ifWritingAuxFile\immediate\write\auxfile{\noexpand\xdef\noexpand#1{#1}}\fi%
}

\def\subsection{\Subsection}

\def\romappno{\uppercase\expandafter{\romannumeral\appno}}
\def\makeNormalizedRomappno{%
\expandafter\makeNormal\expandafter\normalizedromappno%
\expandafter{\romannumeral\appno}%
\edef\normalizedromappno{\uppercase{\normalizedromappno}}}
\def\Appendix#1{\global\advance\appno by1\relax\global\meqno=1\global\secno=0%
\global\subsecno=0%
\bigbreak\bigskip
\centerline{\twelvepoint \bf Appendix %
\romappno. #1}%
\par\nobreak\medskip\nobreak}
\def\tagappendix#1{\makeNormalizedRomappno%
\warnIfChanged#1{\normalizedromappno}%
\xdef#1{\normalizedromappno}%
\ifWritingAuxFile\immediate\write\auxfile{\noexpand\xdef\noexpand#1{#1}}\fi%
}
\def\appendix{\Appendix}
\def\Subappendix#1{\global\advance\subsecno by1\relax\medskip %
\leftline{\bf\romappno.\the\subsecno\ #1}%
\par\nobreak\smallskip\nobreak}
\def\tagsubappendix#1{\makeNormalizedRomappno%
\warnIfChanged#1{\normalizedromappno.\the\subsecno}%
\xdef#1{\normalizedromappno.\the\subsecno}%
\ifWritingAuxFile\immediate\write\auxfile{\noexpand\xdef\noexpand#1{#1}}\fi%
}

\def\eqn#1{\makeNormalizedRomappno%
\ifnum\secno>0%
  \warnIfChanged#1{\the\secno.\the\meqno}%
  \eqno(\the\secno.\the\meqno)\xdef#1{\the\secno.\the\meqno}%
     \global\advance\meqno by1
\else\ifnum\appno>0%
  \warnIfChanged#1{\normalizedromappno.\the\meqno}%
  \eqno({\rm\romappno}.\the\meqno)%
      \xdef#1{\normalizedromappno.\the\meqno}%
     \global\advance\meqno by1
\else%
  \warnIfChanged#1{\the\meqno}%
  \eqno(\the\meqno)\xdef#1{\the\meqno}%
     \global\advance\meqno by1
\fi\fi%
\eqlabeL#1%
\ifWritingAuxFile\immediate\write\auxfile{\noexpand\xdef\noexpand#1{#1}}\fi%
}
\def\defeqn#1{\makeNormalizedRomappno%
\ifnum\secno>0%
  \warnIfChanged#1{\the\secno.\the\meqno}%
  \xdef#1{\the\secno.\the\meqno}%
     \global\advance\meqno by1
\else\ifnum\appno>0%
  \warnIfChanged#1{\normalizedromappno.\the\meqno}%
  \xdef#1{\normalizedromappno.\the\meqno}%
     \global\advance\meqno by1
\else%
  \warnIfChanged#1{\the\meqno}%
  \xdef#1{\the\meqno}%
     \global\advance\meqno by1
\fi\fi%
\eqlabeL#1%
\ifWritingAuxFile\immediate\write\auxfile{\noexpand\xdef\noexpand#1{#1}}\fi%
}
\def\anoneqn{\makeNormalizedRomappno%
\ifnum\secno>0
  \eqno(\the\secno.\the\meqno)%
     \global\advance\meqno by1
\else\ifnum\appno>0
  \eqno({\rm\normalizedromappno}.\the\meqno)%
     \global\advance\meqno by1
\else
  \eqno(\the\meqno)%
     \global\advance\meqno by1
\fi\fi%
}
\def\mfig#1#2{\ifx#20
\else\global\advance\figno by1%
\relax#1\the\figno%
\warnIfChanged#2{\the\figno}%
\xdef#2{\the\figno}%
\reflabeL#2%
\ifWritingAuxFile\immediate\write\auxfile{\noexpand\xdef\noexpand#2{#2}}\fi\fi%
}

\catcode`@=11 

\newif\ifFiguresInText\FiguresInTexttrue
\newif\if@FigureFileCreated
\newwrite\capfile
\newwrite\figfile

\newif\ifcaption
\captiontrue
\def\captionsize{\tenrm}
\def\PlaceTextFigure#1#2#3#4{%
\vskip 0.5truein%
#3\hfil\epsfbox{#4}\hfil\break%
\ifcaption\vskip 5pt\hfil\vbox{\captionsize Figure #1. #2}\hfil\fi%
\vskip10pt}
\def\PlaceEndFigure#1#2{%
\epsfxsize=\hsize\epsfbox{#2}\vfill\centerline{Figure #1.}\eject}

\def\LoadFigure#1#2#3#4{%
\ifundef#1{\phantom{\mfig{}#1}}\else
\ifx#10
\else
\ifWritingAuxFile\immediate\write\auxfile{\noexpand\xdef\noexpand#1{#1}}\fi\fi
\fi%
\ifFiguresInText
\PlaceTextFigure{#1}{#2}{#3}{#4}%
\else
\if@FigureFileCreated\else%
\immediate\openout\capfile=\jobname.caps%
\immediate\openout\figfile=\jobname.figs%
@FigureFileCreatedtrue\fi%
\immediate\write\capfile{\noexpand\item{Figure \noexpand#1.\ }{#2}\vskip10pt}%
\immediate\write\figfile{\noexpand\PlaceEndFigure\noexpand#1{\noexpand#4}}%
\fi}

\def\listfigs{\ifFiguresInText\else%
\vfill\eject\immediate\closeout\capfile
\immediate\closeout\figfile%
\centerline{{\bf Figures}}\bigskip\frenchspacing%
\catcode`@=11 
\def\captionsize{\tenrm}
\input \jobname.caps\vfill\eject\nonfrenchspacing%
\catcode`\@=\active
\catcode`@=12  
\input\jobname.figs\fi}

\font\ninerm=cmr9
\font\eightrm=cmr8
\font\sixrm=cmr6

\def\loadtrueseventeenpoint{
 \font\seventeenrm=cmr10 at 17.28truept
 \font\seventeeni=cmmi10 at 17.28truept
 \font\seventeenbf=cmbx10 at 17.28truept
 \font\seventeenit=cmti10 at 17.28truept
 \font\seventeensl=cmsl10 at 17.28truept
 \font\seventeensy=cmsy10 at 17.28truept
}
\def\loadfourteenpoint{
\font\fourteenrm=cmr10 at 14.4pt
\font\fourteeni=cmmi10 at 14.4pt
\font\fourteenit=cmti10 at 14.4pt
\font\fourteensl=cmsl10 at 14.4pt
\font\fourteensy=cmsy10 at 14.4pt
\font\fourteenbf=cmbx10 at 14.4pt
}
\def\loadtruetwelvepoint{
\font\twelverm=cmr10 at 12truept
\font\twelvei=cmmi10 at 12truept
\font\twelveit=cmti10 at 12truept
\font\twelvesl=cmsl10 at 12truept
\font\twelvesy=cmsy10 at 12truept
\font\twelvebf=cmbx10 at 12truept
\font\twelvesc=cmcsc10 at 12truept
}

\font\ninei=cmmi9
\font\eighti=cmmi8
\font\sixi=cmmi6
\skewchar\ninei='177 \skewchar\eighti='177 \skewchar\sixi='177

\font\ninesy=cmsy9
\font\eightsy=cmsy8
\font\sixsy=cmsy6
\skewchar\ninesy='60 \skewchar\eightsy='60 \skewchar\sixsy='60

\font\ninebf=cmbx9
\font\eightbf=cmbx8
\font\sixbf=cmbx6

\font\ninett=cmtt9
\font\eighttt=cmtt8

\hyphenchar\tentt=-1 
\hyphenchar\ninett=-1
\hyphenchar\eighttt=-1         

\font\ninesl=cmsl9
\font\eightsl=cmsl8

\font\nineit=cmti9
\font\eightit=cmti8
\font\sevenit=cmti7

\scriptfont\itfam=\sevenit

                      
\newskip\ttglue
\def\tenpoint{\def\rm{\fam0\tenrm}%
  \textfont0=\tenrm \scriptfont0=\sevenrm \scriptscriptfont0=\fiverm
  \textfont1=\teni \scriptfont1=\seveni \scriptscriptfont1=\fivei
  \textfont2=\tensy \scriptfont2=\sevensy \scriptscriptfont2=\fivesy
  \textfont3=\tenex \scriptfont3=\tenex \scriptscriptfont3=\tenex
  \def\it{\fam\itfam\tenit}%
      \textfont\itfam=\tenit\scriptfont\itfam=\sevenit
  \def\sl{\fam\slfam\tensl}\textfont\slfam=\tensl
  \def\bf{\fam\bffam\tenbf}\textfont\bffam=\tenbf \scriptfont\bffam=\sevenbf
  \scriptscriptfont\bffam=\fivebf
  \normalbaselineskip=12pt
  \let\sc=\eightrm
  \let\big=\tenbig
  \setbox\strutbox=\hbox{\vrule height8.5pt depth3.5pt width\z@}%
  \normalbaselines\rm}

\def\twelvepoint{\def\rm{\fam0\twelverm}%
  \textfont0=\twelverm \scriptfont0=\ninerm \scriptscriptfont0=\sevenrm
  \textfont1=\twelvei \scriptfont1=\ninei \scriptscriptfont1=\seveni
  \textfont2=\twelvesy \scriptfont2=\ninesy \scriptscriptfont2=\sevensy
  \textfont3=\tenex \scriptfont3=\tenex \scriptscriptfont3=\tenex
  \def\it{\fam\itfam\twelveit}\textfont\itfam=\twelveit
  \def\sl{\fam\slfam\twelvesl}\textfont\slfam=\twelvesl
  \def\bf{\fam\bffam\twelvebf}\textfont\bffam=\twelvebf%
  \scriptfont\bffam=\ninebf
  \scriptscriptfont\bffam=\sevenbf
  \normalbaselineskip=12pt
  \let\sc=\eightrm
  \let\big=\tenbig
  \setbox\strutbox=\hbox{\vrule height8.5pt depth3.5pt width\z@}%
  \normalbaselines\rm}

\def\fourteenpoint{\def\rm{\fam0\fourteenrm}%
  \textfont0=\fourteenrm \scriptfont0=\tenrm \scriptscriptfont0=\sevenrm
  \textfont1=\fourteeni \scriptfont1=\teni \scriptscriptfont1=\seveni
  \textfont2=\fourteensy \scriptfont2=\tensy \scriptscriptfont2=\sevensy
  \textfont3=\tenex \scriptfont3=\tenex \scriptscriptfont3=\tenex
  \def\it{\fam\itfam\fourteenit}\textfont\itfam=\fourteenit
  \def\sl{\fam\slfam\fourteensl}\textfont\slfam=\fourteensl
  \def\bf{\fam\bffam\fourteenbf}\textfont\bffam=\fourteenbf%
  \scriptfont\bffam=\tenbf
  \scriptscriptfont\bffam=\sevenbf
  \normalbaselineskip=17pt
  \let\sc=\elevenrm
  \let\big=\tenbig                                          
  \setbox\strutbox=\hbox{\vrule height8.5pt depth3.5pt width\z@}%
  \normalbaselines\rm}

\def\seventeenpoint{\def\rm{\fam0\seventeenrm}%
  \textfont0=\seventeenrm \scriptfont0=\fourteenrm \scriptscriptfont0=\tenrm
  \textfont1=\seventeeni \scriptfont1=\fourteeni \scriptscriptfont1=\teni
  \textfont2=\seventeensy \scriptfont2=\fourteensy \scriptscriptfont2=\tensy
  \textfont3=\tenex \scriptfont3=\tenex \scriptscriptfont3=\tenex
  \def\it{\fam\itfam\seventeenit}\textfont\itfam=\seventeenit
  \def\sl{\fam\slfam\seventeensl}\textfont\slfam=\seventeensl
  \def\bf{\fam\bffam\seventeenbf}\textfont\bffam=\seventeenbf%
  \scriptfont\bffam=\fourteenbf
  \scriptscriptfont\bffam=\twelvebf
  \normalbaselineskip=21pt
  \let\sc=\fourteenrm
  \let\big=\tenbig                                          
  \setbox\strutbox=\hbox{\vrule height 12pt depth 6pt width\z@}%
  \normalbaselines\rm}

\def\ninepoint{\def\rm{\fam0\ninerm}%
  \textfont0=\ninerm \scriptfont0=\sixrm \scriptscriptfont0=\fiverm
  \textfont1=\ninei \scriptfont1=\sixi \scriptscriptfont1=\fivei
  \textfont2=\ninesy \scriptfont2=\sixsy \scriptscriptfont2=\fivesy
  \textfont3=\tenex \scriptfont3=\tenex \scriptscriptfont3=\tenex
  \def\it{\fam\itfam\nineit}\textfont\itfam=\nineit
  \def\sl{\fam\slfam\ninesl}\textfont\slfam=\ninesl
  \def\bf{\fam\bffam\ninebf}\textfont\bffam=\ninebf \scriptfont\bffam=\sixbf
  \scriptscriptfont\bffam=\fivebf
  \normalbaselineskip=11pt
  \let\sc=\sevenrm
  \let\big=\ninebig
  \setbox\strutbox=\hbox{\vrule height8pt depth3pt width\z@}%
  \normalbaselines\rm}

\def\eightpoint{\def\rm{\fam0\eightrm}%
  \textfont0=\eightrm \scriptfont0=\sixrm \scriptscriptfont0=\fiverm%
  \textfont1=\eighti \scriptfont1=\sixi \scriptscriptfont1=\fivei%
  \textfont2=\eightsy \scriptfont2=\sixsy \scriptscriptfont2=\fivesy%
  \textfont3=\tenex \scriptfont3=\tenex \scriptscriptfont3=\tenex%
  \def\it{\fam\itfam\eightit}\textfont\itfam=\eightit%
  \def\sl{\fam\slfam\eightsl}\textfont\slfam=\eightsl%
  \def\bf{\fam\bffam\eightbf}\textfont\bffam=\eightbf \scriptfont\bffam=\sixbf%
  \scriptscriptfont\bffam=\fivebf%
  \normalbaselineskip=9pt%
  \let\sc=\sixrm%
  \let\big=\eightbig%
  \setbox\strutbox=\hbox{\vrule height7pt depth2pt width\z@}%
  \normalbaselines\rm}
  \let\sc=\eightrm

\def\tenbig#1{{\hbox{$\left#1\vbox to8.5pt{}\right.\n@space$}}}
\def\ninebig#1{{\hbox{$\textfont0=\tenrm\textfont2=\tensy
  \left#1\vbox to7.25pt{}\right.\n@space$}}}
\def\eightbig#1{{\hbox{$\textfont0=\ninerm\textfont2=\ninesy
  \left#1\vbox to6.5pt{}\right.\n@space$}}}

\def\footnote#1{\edef\@sf{\spacefactor\the\spacefactor}#1\@sf
      \insert\footins\bgroup\eightpoint
      \interlinepenalty100 \let\par=\endgraf
        \leftskip=\z@skip \rightskip=\z@skip
        \splittopskip=10pt plus 1pt minus 1pt \floatingpenalty=20000
        \smallskip\item{#1}\bgroup\strut\aftergroup\@foot\let\next}
\skip\footins=12pt plus 2pt minus 4pt 
\dimen\footins=30pc 

\newinsert\margin
\dimen\margin=\maxdimen
\def\titlefont{\seventeenpoint}
\loadtruetwelvepoint 
\loadtrueseventeenpoint

\def\eatOne#1{}
\def\ifundef#1{\expandafter\ifx%
\csname\expandafter\eatOne\string#1\endcsname\relax}
\def\notTrue{\iffalse}\def\isTrue{\iftrue}
\def\ifdef#1{{\ifundef#1%
\aftergroup\notTrue\else\aftergroup\isTrue\fi}}
\def\use#1{\ifundef#1\linemessage{Warning: \string#1 is undefined.}%
{\tt \string#1}\else#1\fi}



%
\catcode`"=11
\let\quote="
\catcode`"=12
\chardef\foo="22
\global\newcount\refno \global\refno=1
\newwrite\rfile
\newlinechar=`\^^J
\def\@ref#1#2{\the\refno\n@ref#1{#2}}
\def\h@ref#1#2#3{\href{#3}{\the\refno}\n@ref#1{#2}}
\def\n@ref#1#2{\xdef#1{\the\refno}%
\ifnum\refno=1\immediate\openout\rfile=\jobname.refs\fi%
\immediate\write\rfile{\noexpand\item{[\noexpand#1]\ }#2.}%
\global\advance\refno by1}
\def\nref{\n@ref} 
\def\ref{\@ref}   
\def\hrref{\h@ref}
\def\lref#1#2{\the\refno\xdef#1{\the\refno}%
\ifnum\refno=1\immediate\openout\rfile=\jobname.refs\fi%
\immediate\write\rfile{\noexpand\item{[\noexpand#1]\ }#2\semi}%
\global\advance\refno by1}
\def\cref#1{\immediate\write\rfile{#1\semi}}

\def\preref#1#2{\gdef#1{\@ref#1{#2}}}

\def\semi{;\hfil\noexpand\break}

\def\listrefs{\vfill\eject\immediate\closeout\rfile
\centerline{{\bf References}}\bigskip\frenchspacing%
\input \jobname.refs\vfill\eject\nonfrenchspacing}

\def\inputAuxIfPresent#1{\immediate\openin1=#1
\ifeof1\message{No file \auxfileName; I'll create one.
}\else\closein1\relax\input\auxfileName\fi%
}
\def\NPB{Nucl.\ Phys.\ B}




\newif\ifWritingAuxFile
\newwrite\auxfile
\def\SetUpAuxFile{%
\xdef\auxfileName{\jobname.aux}%
\inputAuxIfPresent{\auxfileName}%
\WritingAuxFiletrue%
\immediate\openout\auxfile=\auxfileName}

\def\L{\left(}\def\R{\right)}
\def\LP{\left.}\def\RP{\right.}
\def\LB{\left[}\def\RB{\right]}

\def\RV{\right|}

\def\bye{\par\vfill\supereject%
\ifAnyCounterChanged\linemessage{
Some counters have changed.  Re-run tex to fix them up.}\fi%
\end}

\catcode`\@=\active
\catcode`@=12  
\catcode`\"=\active

 \def\Tr{\mathop{\rm Tr}\nolimits}
 \def\pol{\varepsilon}

\def\Tr{\mathop{\rm Tr}\nolimits}

\def\pol{\varepsilon}

\def\ksl{\slashed{k}}

\def\L{\left(}\def\R{\right)}
\def\LP{\left.}\def\RP{\right.}
\def\spa#1.#2{\left\langle#1\,#2\right\rangle}
\def\spb#1.#2{\left[#1\,#2\right]}
\def\lor#1.#2{\left(#1\,#2\right)}
\def\sand#1.#2.#3{%
\left\langle\smash{#1}{\vphantom1}^{-}\right|{#2}%
\left|\smash{#3}{\vphantom1}^{-}\right\rangle}
\def\sandp#1.#2.#3{%
\left\langle\smash{#1}{\vphantom1}^{-}\right|{#2}%
\left|\smash{#3}{\vphantom1}^{+}\right\rangle}
\def\sandpp#1.#2.#3{%
\left\langle\smash{#1}{\vphantom1}^{+}\right|{#2}%
\left|\smash{#3}{\vphantom1}^{+}\right\rangle}
\def\sandpm#1.#2.#3{%
\left\langle\smash{#1}{\vphantom1}^{+}\right|{#2}%
\left|\smash{#3}{\vphantom1}^{-}\right\rangle}
\def\sandmp#1.#2.#3{%
   \left\langle\smash{#1}{\vphantom1}^{-}\right|{#2}%
    \left|\smash{#3}{\vphantom1}^{+}\right\rangle}
\catcode`@=11  
\def\meqalign#1{\,\vcenter{\openup1\jot\m@th
   \ialign{\strut\hfil$\displaystyle{##}$ && $\displaystyle{{}##}$\hfil
             \crcr#1\crcr}}\,}
\catcode`@=12  

\SetUpAuxFile
\hfuzz 20pt
\overfullrule 0pt

\newcount\oldpage

\def\peterc{\ifnum\pageno>\oldpage
\headline{
}\fi
\rm
}

\def\e{\epsilon}
\def\tree{{\rm tree\vphantom{p}}}
\def\oneloop{{\rm 1\hbox{\sevenrm-}loop}}

\def\Split{\mathop{\rm Split}\nolimits}
\def\Soft{\mathop{\rm Soft}\nolimits}

\def\Ctree{\Split^\tree}
\def\Cone{\Split^\oneloop}

\def\tcdot{\mskip -1mu\cdot\mskip-1mu}
\def\ll#1{{\lambda_{#1}}}
\def\la{\ll{a}}
\def\lb{\ll{b}}
\def\lc{\ll{c}}
\def\ah{{\hat a}}
\def\bh{{\hat b}}

\def\ls{\ll{s}}
\def\llongrightarrow{%
\relbar\mskip-0.5mu\joinrel\mskip-0.5mu\relbar\mskip-0.5mu\joinrel\longrightarrow}
\def\inlimit^#1{\buildrel#1\over\llongrightarrow}

\def\phpol{{\rm ph.\ pol.}}
\def\Ord{{\cal O}}
\def\ksl{\slashed{k}}
\def\F#1#2{\,{{\vphantom{F}}_{#1}F_{#2}}}
\def\Li{\mathop{\rm Li}\nolimits}

\loadfourteenpoint

\input epsf

\noindent\nopagenumbers
hep-ph/9903515 \hfill{Saclay/SPhT--T99/032}

\leftlabelstrue
\vskip -1.0 in
\Title{One-Loop Splitting Amplitudes in Gauge Theory}

\baselineskip17truept
\centerline{David A. Kosower and Peter Uwer}
\baselineskip12truept
\centerline{\it Service de Physique Th\'eorique${}^{\natural}$}
\centerline{\it Centre d'Etudes de Saclay}
\centerline{\it F-91191 Gif-sur-Yvette cedex, France}
\centerline{\tt kosower@spht.saclay.cea.fr}
\centerline{\tt uwer@spht.saclay.cea.fr}

\vskip 0.2in\baselineskip13truept

\vskip 0.5truein
\centerline{\bf Abstract}

{\narrower 

We recompute the functions
describing the collinear factorization of
one-loop amplitudes using the unitarity-based
method.  We present the results in a form
suitable for use as an ingredient in
two-loop calculations.  We also present
a function summarizing the behavior
at one loop in both the soft and collinear
limits.

}
\vskip 0.3truein


\vfill
\vskip 0.1in
\noindent\hrule width 3.6in\hfil\break
\noindent
${}^{\natural}$Laboratory of the
{\it Direction des Sciences de la Mati\`ere\/}
of the {\it Commissariat \`a l'Energie Atomique\/} of France.\hfil\break

\Date{}

\line{}

\baselineskip17pt
%

\preref\Color{%
F. A. Berends and W. T. Giele,
Nucl.\ Phys.\ B294:700 (1987)\semi
D. A.\ Kosower, B.-H.\ Lee and V. P. Nair, Phys.\ Lett.\ 201B:85 (1988)\semi
M.\ Mangano, S. Parke and Z.\ Xu, Nucl.\ Phys.\ B298:653 (1988)\semi
Z. Bern and D. A.\ Kosower, Nucl.\ Phys.\ B362:389 (1991)}

\preref\GG{W.T.\ Giele and E.W.N.\ Glover,
Phys.\ Rev.\ D46:1980 (1992)}
\preref\GGK{W.T.\ Giele, E. W. N.\ Glover and D. A. Kosower,
Nucl.\ Phys.\ B403:633 (1993) [hep-ph/9302225]}

\preref\Recurrence{F. A. Berends and W. T. Giele, Nucl.\ Phys.\ B306:759 (1988)\semi
D. A. Kosower, Nucl.\ Phys.\ B335:23 (1990)}

\preref\ManganoReview{%
M. Mangano and S.J. Parke, Phys.\ Rep.\ 200:301 (1991)}

\preref\CS{%
S. Catani and M. Seymour, Phys.\ Lett.\ B378:287 (1996) [hep-ph/9602277];
Nucl.\ Phys.\ B485:291 (1997) [hep-ph/9605323] (err. ibid. B510:503 (1997))}

\preref\CST{S. Catani, M. H. Seymour, and Z. Tr\'ocs\'anyi,
       Phys.\ Rev.\ D55:6819 (1997) [hep-ph/9610553]}

\preref\SoftGluonReview{A. Bassetto, M. Ciafaloni, and G. Marchesini, 
 Phys.\ Rep.\ 100:201 (1983)}
\preref\AP{G. Altarelli and G. Parisi, Nucl.\ Phys.\ B126:298 (1977)}
\preref\FKS{S. Frixione, Z. Kunszt, and A. Signer, 
   Nucl.\ Phys.\ B467:399 (1996) [hep-ph/9512328]}

\preref\Byckling{E. Byckling and K. Kajantie, {\it Particle Kinematics\/}
(Wiley, 1973)}

\preref\KobaNielsenR{Z. Koba and H. B. Nielsen, Nucl.\ Phys.\ B10:633 (1969)\semi
J. H. Schwarz, Phys.\ Rep.\ 89:223 (1982)}

\preref\CDR{J. C.\ Collins, {\it Renormalization} (Cambridge University Press, 1984)}
\preref\HV{G. 't Hooft and M. Veltman, Nucl.\ Phys.\ B44:189 (1972)}
\preref\DimRed{W. Siegel, Phys.\ Lett.\ 84B:193 (1979)\semi
D.M.\ Capper, D.R.T.\ Jones and P. van Nieuwenhuizen, Nucl.\ Phys.\
B167:479 (1980)\semi
L.V.\ Avdeev and A.A.\ Vladimirov, Nucl.\ Phys.\ B219:262 (1983)}
\preref\FDHS{Z. Bern and D. A.\ Kosower, \NPB 379:451 (1992)}

\preref\BernChalmers{Z. Bern and G. Chalmers, Nucl.\ Phys.\ B447:465 (1995) 
  [hep-ph/9503236]}
\preref\Pentagon{Z. Bern, L. Dixon, and D. A. Kosower,
            Nucl.\ Phys.\ B412:751 (1994) [hep-ph/9306240]}

\preref\SusyOne{Z. Bern, L. Dixon, D. C. Dunbar and D. A. Kosower, 
        Nucl.\ Phys.\ B435:59 (1995) [hep-ph/9409265]}
\preref\SusyFour{Z. Bern, L. Dixon, D. C. Dunbar, and D. A. Kosower,
Nucl.\ Phys.\ B425:217 (1994) [hep-ph/9403226]}
\preref\qqggg{Z. Bern, L. Dixon, and D. A. Kosower,
Nucl.\ Phys.\  B437:259 (1995) [hep-ph/9409393]}
\preref\AnnRev{Z. Bern, L. Dixon, and D. A. Kosower,
Ann.\ Rev.\ Nucl.\ Part.\ Sci.\ 46 109 (1996) [hep-ph/9602280]}

\preref\EnglishDoubleCollinear{J. M. Campbell and E. W. N. Glover,
Nucl.\ Phys.\ B527:264 (1998) [hep-ph/9710255]}
\preref\ItalianDoubleCollinear{S. Catani and M. Grazzini, preprint hep-ph/9810389}

\preref\BDS{Z. Bern, V. Del Duca, and C. R. Schmidt, 
   Phys.\ Lett.\ B445:168 (1998) [hep-ph/9810409]}
\preref\AllOrders{D. A. Kosower, hep-ph/9901201}

\preref\BrownFeynman{L. M. Brown and R. P. Feynman, Phys.\ Rev.\ 85:231 (1952)}
\preref\Passarino{G. Passarino and M. Veltman, \NPB{160:151 (1979)}}

\preref\SpinorHelicity{
F.\ A.\ Berends, R.\ Kleiss, P.\ De Causmaecker, R.\ Gastmans and T.\ T.\ Wu,
        Phys.\ Lett.\ 103B:124 (1981)\semi
P.\ De Causmaeker, R.\ Gastmans,  W.\ Troost and  T.\ T.\ Wu,
Nucl. Phys. B206:53 (1982)\semi
R.\ Kleiss and W.\ J.\ Stirling,
   Nucl.\ Phys.\ B262:235 (1985)\semi
   J.\ F.\ Gunion and Z.\ Kunszt, Phys.\ Lett.\ 161B:333 (1985)\semi
 R.\ Gastmans and T. T.\ Wu,
{\it The Ubiquitous Photon: Helicity Method for QED and QCD} 
(Clarendon Press,1990)}
\preref\XZC{
Z. Xu, D.-H.\ Zhang and L. Chang, Nucl.\ Phys.\ B291:392 (1987)}

\preref\Antenna{D. A. Kosower, Phys.\ Rev.\ D57:5410 (1998) [hep-ph/9710213]}

\preref\SWI{M. T. Grisaru, H. N. Pendleton and P. van
Nieuwenhuizen,
Phys. Rev. D15:996 (1977)\semi
M. T. Grisaru and H.N. Pendleton, Nucl. Phys. B124:81 (1977)}
\preref\UseSWI{
S. J. Parke and T. Taylor, Phys. Lett. B157:81 (1985)\semi
Z. Kunszt, Nucl. Phys. B271:333 (1986)}
\preref\Lewin{L. Lewin, {\it Polylogarithms and associated functions\/}
(North-Holland,1981)}

\preref\ParkeTaylor{S. J. Parke and T. R. Taylor,
Phys.\ Rev.\ Lett.\ 56:2459 (1986)}

\preref\GR{I. S. Gradshteyn and I. M. Ryzhik, {\it Table of Integrals,
Series, and Products\/}, ed.\ A. Jeffrey 
(Academic Press, 1980)}

\preref\BDKS{Z. Bern, V. Del Duca, W. B. Kilgore, and C. R. Schmidt, in
preparation}

\section{Introduction}
\vskip 10pt

The properties of non-Abelian gauge-theory amplitudes in soft
and collinear limits play an important role
both in formal proofs and in explicit calculations.
The singularities that arise in these limits create technical
obstacles to calculations of amplitudes and cross
sections.  Their universal structure, however,
provides a means of resolving these
problems.  The universality of the limits is furthermore essential to the
applicability of quantum chromodynamics to 
short-distance physics.  
It also provides a practical means for checking calculations of
higher-point amplitudes.
Explicit calculations of the
functions governing these limits thus play an important
role in the program of precision quantum chromodynamical
calculations.

In these singular limits, on-shell amplitudes factorize
into sums of lower-point amplitudes multiplied by
universal functions.  In the collinear limit, these
functions are called {\it splitting amplitudes\/}.
At tree level, one may derive the splitting amplitudes from
a string representation~[\use\ManganoReview] or from
the Berends--Giele recurrence relations~[\Recurrence].
The squares of these tree-level functions, 
summed over helicities,
yield the Altarelli--Parisi kernels~[\use\AP,\use\SoftGluonReview].

Beyond leading order, the splitting amplitudes do not
yield the Altarelli--Parisi kernel directly, and
the relation between the two quantities remains to be clarified.
The explicit forms of the splitting amplitudes
at one-loop
through $\Ord(\e^0)$ (in dimensional
regularization, with $D=4-2\e$) have previously been extracted by
taking the collinear limit explicitly in various five-point
amplitudes~[\use\SusyFour,\use\qqggg], or from an analysis of
one-loop integrals~[\use\BernChalmers].  
In addition, Bern, Del~Duca, and Schmidt~[\use\BDS]
 have recently given an expression for the gluon splitting
amplitude to all orders in the dimensional regulator $\e$.

Many of the complicated one-loop computations of recent years
have been performed using the unitarity-based method
developed by Bern, Dixon, Dunbar, and one of the 
authors~[\use\SusyOne,\use\AnnRev].  The method also leads
to a concise proof~[\use\AllOrders] of collinear factorization to 
all loop orders in gauge theories, and as a by-product
provides a compact formula for computing the 
splitting amplitudes explicitly.
In this paper, we use this formula to compute the 
various one-loop splitting amplitudes, thus
providing complete forms to all orders in $\e$ for both
gluon and fermion external states.
The splitting amplitudes presented here will be useful
in computations of physical observables such as 
jet cross-sections, at next-to-next-to-leading order.
In addition to the one-loop splitting amplitudes,
for such purposes one would also need the tree-level functions describing
factorization when three particles become collinear,
computed by Campbell and Glover~[\use\EnglishDoubleCollinear]
 and by Catani and Grazzini~[\use\ItalianDoubleCollinear].

The paper is organized as follows.
We review the structure of collinear factorization
in the next section.
We then present a detailed 
example calculation, the gluon splitting amplitude with
selected external helicities, in section~\use\ExampleSection.
We present the complete set of gluon and fermion splitting amplitudes
in section~\use\GeneralSection, and discuss various checks
and comparisons in section~\use\CheckSection.  In 
section~\use\AntennaSection, we use these splitting amplitudes
to obtain a function which interpolates smoothly between the
collinear and soft limits.  

\section{The Structure of Collinear Limits}
\vskip 10pt

The properties of gauge theories are easiest to discuss
in the context of a color decomposition~[\use\Color].
At tree level, for all-gluon amplitudes
such a decomposition takes the form,
$$
{\cal A}_n^\tree(\{k_i,\lambda_i,a_i\}) = 
g^{n-2} \sum_{\sigma \in S_n/Z_n} \Tr(T^{a_{\sigma(1)}}\cdots T^{a_{\sigma(n)}})\,
A_n^\tree(\sigma(1^{\lambda_1},\ldots,n^{\lambda_n}))\,,
\eqn\TreeColorDecomposition$$
where $S_n/Z_n$ is the group of non-cyclic permutations
on $n$ symbols, $j^{\lambda_j}$ denotes the $j$-th momentum
and helicity, and $g$ is the gauge coupling.  
We use the normalization $\Tr(T^a T^b) = \delta^{ab}$ for the generators
of $SU(N)$.
One can write analogous formul\ae\ for amplitudes
with quark-antiquark pairs or uncolored external lines.
The color-ordered or partial amplitude $A_n$ is gauge invariant.

\LoadFigure\TreeFactorizationFigure
{\baselineskip 13 pt
\noindent\narrower\ninerm
A schematic depiction of the collinear factorization of tree-level
amplitudes, with the amplitudes labelled clockwise.
}  {\epsfxsize 4.2 truein}{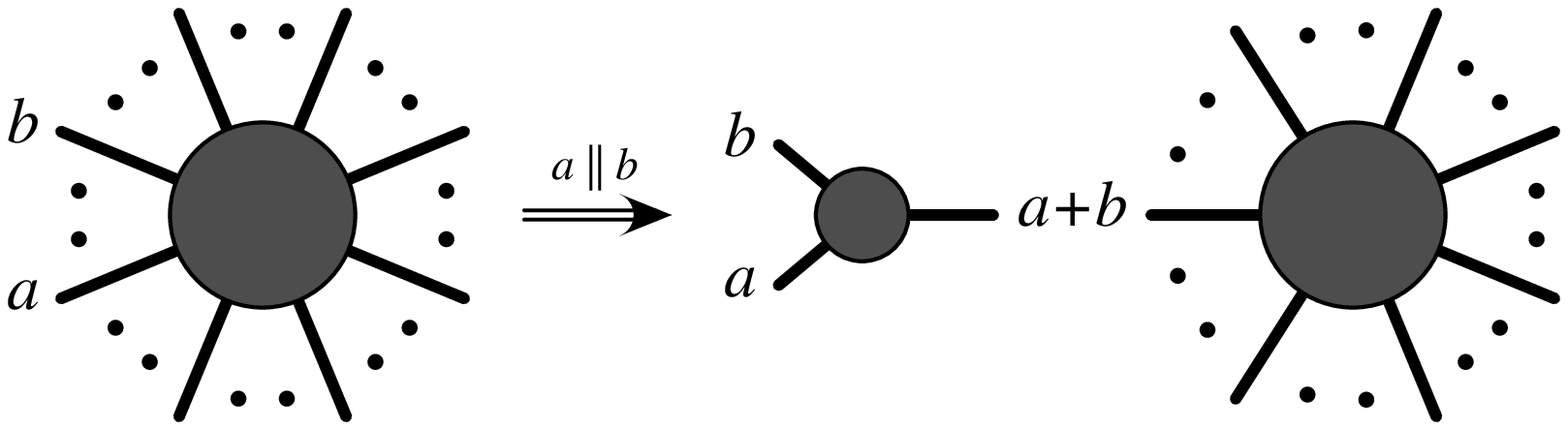}{}

In the collinear limit, $k_a \parallel k_b$ of two adjacent legs,
the color-ordered amplitude $A_n$ is singular.  (It is finite
when the two collinear legs are not adjacent arguments.)  This
singular behavior has a universal form expressed by the 
tree-level factorization
equation,
$$\eqalign{
&A_{n}^\tree(1,\ldots,a^\la,b^\lb,\ldots,n) 
   \inlimit^{k_a \tcdot k_b\rightarrow 0}
 \sum_{\phpol\ \sigma}
 \Ctree_{-\sigma}(a^\la,b^\lb) 
    A_{n-1}^\tree(1,\ldots,(a+b)^\sigma,\ldots,n)\,,\cr
}\eqn\CollinearA$$
dropping terms finite in the limit.  In this equation, 
$\Ctree$ is the usual tree splitting amplitude, 
and the notation `$a+b$' means $k_a+k_b$.  The
notation `$\phpol$' indicates a sum over physical polarizations only.
(`Physical' here is in the sense of `transverse', and their number 
may depend
on the number of dimensions and on the variant of dimensional
regularization employed.)
This factorization is
depicted schematically in fig.~\use\TreeFactorizationFigure.
It is characteristic of
gauge theories that the splitting amplitude has a square-root singularity,
$\Split \sim 1/\sqrt{s_{ab}}$, rather than a full inverse power of the
two-particle invariant.  

\def\Gr{{\rm Gr}}
At one loop, the analogous color decomposition to~(\use\TreeColorDecomposition) is
$$
{\cal A}_n\L \{k_i,\lambda_i,a_i\}\R =
g^n \sum_J n_J
  \sum_{c=1}^{\lfloor{n/2}\rfloor+1}
      \sum_{\sigma \in S_n/S_{n;c}}
     \Gr_{n;c}\L \sigma \R\,A_{n;c}^{[J]}(\sigma),
\eqn\LoopColorDecomposition$$
where ${\lfloor{x}\rfloor}$ is the largest integer less than or equal to $x$
and $n_J$ is the number of particles of spin $J$.
The leading color-structure factor,
$$
\Gr_{n;1}(1) = N_c\ \Tr\L T^{a_1}\cdots T^{a_n}\R\,,
\anoneqn
$$
is just $N_c$ times the tree color factor, and the subleading color
structures are given by
$$
\Gr_{n;c}(1) = \Tr\L T^{a_1}\cdots T^{a_{c-1}}\R\,
\Tr\L T^{a_c}\cdots T^{a_n}\R.
\anoneqn
$$
$S_n$ is the set of all permutations of $n$ objects,
and $S_{n;c}$ is the subset leaving $\Gr_{n;c}$ invariant.
The decomposition~(\use\LoopColorDecomposition) holds separately
for different spins circulating around the loop.  The usual
normalization conventions take each massless spin-$J$ particle to have two
(helicity) states: gauge bosons, Weyl fermions, and complex scalars.
(For internal particles in the fundamental ($N_c+\bar{N_c}$) representation,
only the single-trace color structure ($c=1$) would be present,
and the corresponding color factor would be smaller by a factor of $N_c$.)

The subleading-color amplitudes $A_{n;c>1}$ are in fact not independent
of the leading-color amplitude $A_{n;1}$.  Rather, they can be expressed as
sums over permutations of the arguments of the latter~[\use\SusyFour].
(For amplitudes with external fermions, the basic objects are primitive
amplitudes~[\use\qqggg] rather than the leading-color one, but the same dependence
of the subleading color amplitudes holds.)

\LoadFigure\OneLoopFactorizationFigure
{\baselineskip 13 pt
\noindent\narrower\ninerm
A schematic depiction of the collinear factorization of one-loop
amplitudes.
}  {\epsfxsize 4.2 truein}{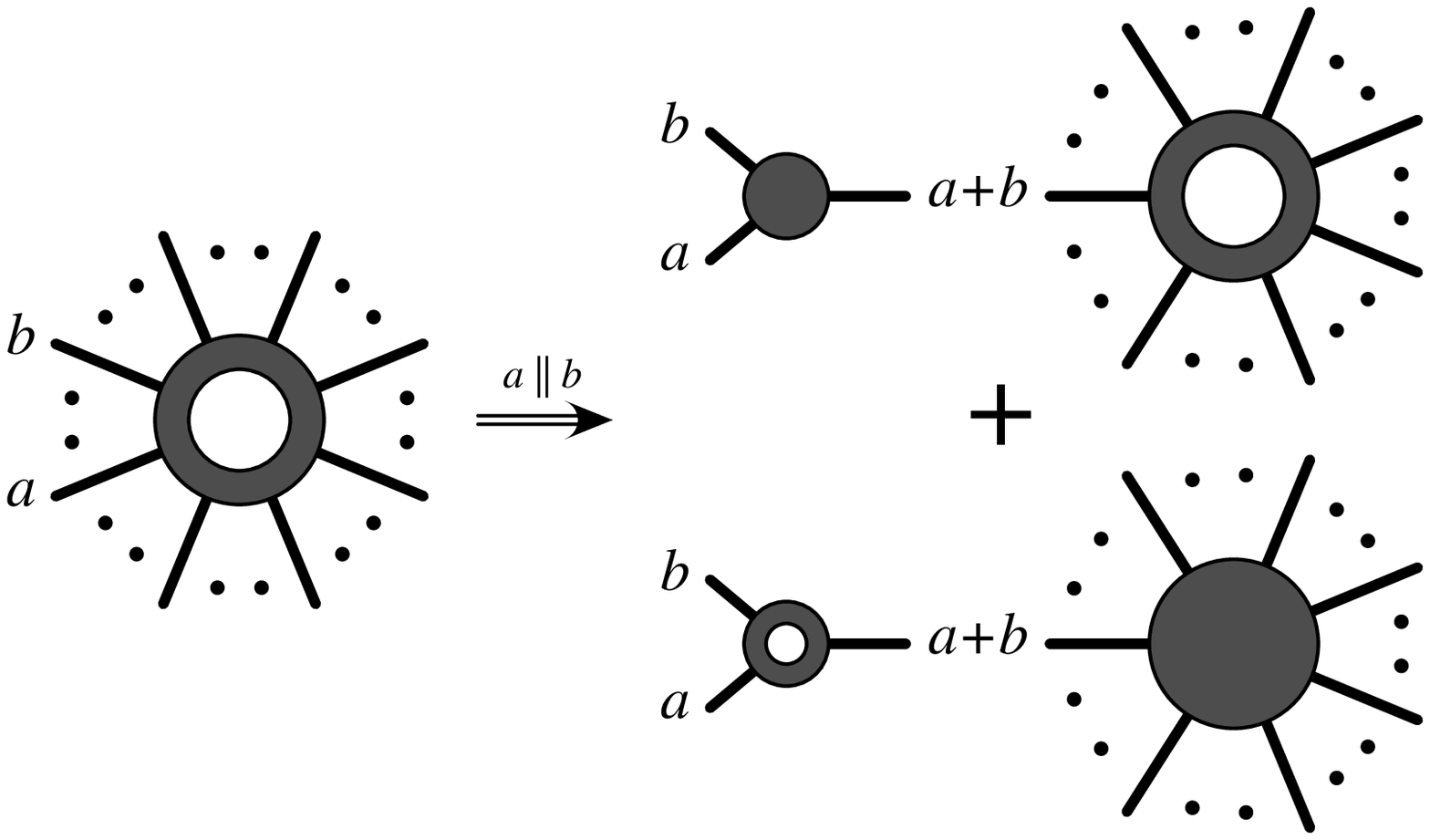}{}
As a result, it suffices to examine the collinear limits of leading-color
amplitudes. The collinear limits of the subleading-color then follow using this
relation.  The leading-color amplitudes obey the following 
factorization~[\use\SusyFour,\use\AllOrders],
$$\eqalign{
 A_n^{\oneloop}&(1,\ldots,a^\la,b^\lb,\ldots,n) 
\;{\buildrel a \parallel b\over{\relbar\mskip-1mu\joinrel\longrightarrow}}\cr
&\sum_{\phpol\ \sigma}  \biggl(
  \Split^\tree_{-\sigma}(a^{\la},b^{\lb})\,
      A_{n-1}^\oneloop(1,\ldots,(a+b)^\sigma,\ldots,n)
\cr &\hskip 20mm  
  +\Split^\oneloop_{-\sigma}(a^\la,b^\lb)\,
      A_{n-1}^\tree(1,\ldots,(a+b)^\sigma,\ldots,n) \biggr) \;.
}\eqn\OneLoopCollinearFactorization$$
This factorization is
depicted schematically in fig.~\use\OneLoopFactorizationFigure.

\LoadFigure\OneLoopSplittingFigure
{\baselineskip 13 pt
\noindent\narrower\ninerm
The defining equation for the one-loop splitting amplitude.
}  {\epsfxsize 4.2 truein}{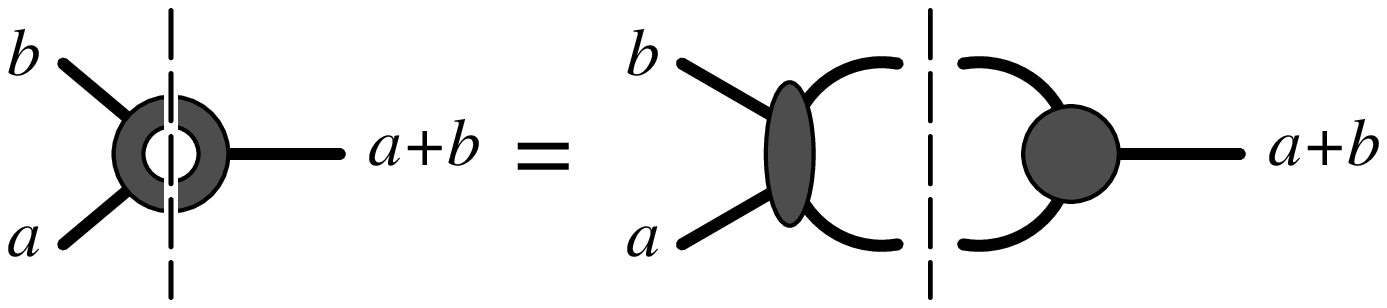}{}
This form was originally deduced from explicit calculations of higher-point
amplitudes, but can also be proven more generally using the unitarity-based
method.  The latter proof also provides an explicit formula for the
one-loop splitting amplitude~[\use\AllOrders],
$$\eqalign{
\Cone_{-\sigma}&(a^\la,b^\lb) = \cr
& 
\sum_{\phpol\ \ll1,\ll2} 
\int {d^{4-2\e}\ell\over (2\pi)^{4-2\e}}\;{i\over\ell^2}
  \Ctree_{-\sigma}((\ell+a+b)^{-\ll2},(-\ell)^{-\ll1}){i\over (\ell+k_a+k_b)^2}
\cr &\hskip 40mm\times
      A_{4}^\tree((-\ell-a-b)^\ll2,a^\la,b^\lb,\ell^\ll1)
}\eqn\MasterFormula$$
The restriction to physical polarizations is important; it will
give rise to transverse projection operators inside the loop.  The
defining equation is depicted graphically in fig.~\use\OneLoopSplittingFigure.

\section{A Sample Calculation}
\tagsection\ExampleSection
\vskip 10pt

\def\polsl{\mathslashed{\pol}}
As an example of a calculation using eqn.~(\use\MasterFormula), we present 
the calculation of the splitting amplitudes $\Cone(a^+,b^-)$ in the
conventional dimensional regularization (CDR) scheme, but restricted to 
physical external helicities.  (The answer in the original 't~Hooft--Veltman (HV)
scheme would be identical.  For a discussion of the differences of the
various schemes, see refs.~[\use\FDHS,\use\CST].)

For this purpose, we need the three-gluon tree-level splitting
amplitude,
$$
\Ctree(P\rightarrow a\,b;z) = 
-{\sqrt2\over s_{ab}} \L  -\pol_a\tcdot\pol_b\,k_b\tcdot\pol_P\, 
+ k_b\tcdot \pol_a\, \pol_P\tcdot\pol_b\,
- k_a\tcdot \pol_b\,\pol_a\tcdot\pol_P\R\,.
\eqn\GluonVertex$$
The variable $z$ denotes the collinear momentum fraction, given 
in the collinear limit by
$$
{q\cdot k_a\over q\cdot (k_a+k_b)}
\anoneqn$$
where $q$ is a {\it reference\/} null momentum which is not collinear
with $k_a$ or $k_b$.  All momenta are taken to be outgoing, $P+k_a+k_b=0$.
The splitting amplitudes will in general have not only an explicit
dependence on the argument $z$, but also an implicit one that arises
from the dependence on the momenta and polarization vectors of legs
$a$ and $b$.  Note that the order of the two arguments $a$ and $b$ is important,
as the definition of $z$ depends on it.
We also need the four-point amplitude $A_4(1^+,2^-,3,4)$,
$$\eqalign{
A_4&(1^+,2^-,3,4) = 
\cr &{i\over 2 s_{14}}
\Bigl[ \sand2.{\polsl_3}.1 \sand2.{\polsl_4}.1
-{\sand2.{\ksl_4}.1 \sandmp2.{\polsl_4\polsl_3}.2\over\spa1.2}
 \cr&\hskip 12mm
+{\sand2.{\ksl_4}.1 \sandpm1.{\polsl_4\polsl_3}.1\over \spb1.2}
-{2 \pol_3\cdot\pol_4\, \sand2.{\ksl_4}.1^2\over s_{12}}
\Bigr]
}\eqn\FourPointPM$$
where the notation indicates that the polarization vectors for the last
two legs are kept generic, and in $D=4-2\e$ dimensions.  
In this formula, $s_{ij} = (k_i+k_j)^2$, and following the notation 
of Xu, Zhang, and Chang [\use\XZC],
$\left|j^\pm\right\rangle$ and $\left\langle j^\pm\right|$ are massless
two-component spinors carrying momentum $k_j$; 
$\spa{j}.l \equiv \left\langle j^-|l^+\right\rangle$ and 
$\spb{j}.l \equiv \left\langle j^+|l^-\right\rangle$
are spinor products~[\use\SpinorHelicity,\use\XZC].

\def\phpol{{\rm ph.\ pol.}}
In the CDR scheme, there are $2-2\e$ `physical' polarizations, reflected in
the identity
$$\eqalign{
\sum_{\phpol\ \sigma} &\pol_\mu^{\sigma} \pol_\nu^{-\sigma} = -g_{\mu\nu}
+ {q_\mu k_\nu + k_\mu q_\nu\over q\cdot k}\,,\cr
}\eqn\PolarizationSum$$
where $k$ is the momentum of the gluon and $q$ is the above-mentioned
reference momentum.

Using this identity, along with
collinear identities such as $\sand{b}.{q}.{a} = -\sqrt{z (1-z)} 2q\cdot P$,
commuting gamma matrices appropriately, and dropping terms which are
finite in the collinear limit,
we find for the integrand of eqn.~(\use\MasterFormula) a sum of
terms,
$$\eqalign{
\Bigl[T_0(\ell) - T_0(P-\ell)
+T_1(\ell) - T_1(P-\ell)
+T_2(\ell)\Bigr] \,{1\over\ell^2 (\ell+k_a+k_b)^2}\,,
}\anoneqn$$
where
\def\psl{\slashed{p}}

$$\eqalign{
T_0(p) &= {1\over \sqrt{2}}{\sqrt{z (1-z)}\, q\cdot P \sand{b}.{\polsl_P}.{a}}
                   {1\over q\cdot p\, s_{a\ell}}\,,\cr
T_1(p) &= {\sqrt{2 z (1-z)}\,q\cdot P} {p\cdot P\over q\cdot p\, s_{{a}\ell}}
                - {q\cdot P \sand{b}.{\polsl_P}.{a}\over\sqrt2 s_{ab}}
                  \, {\sand{b}.{\psl}.{a}\over q\cdot p\, s_{{a}\ell}}\,,\cr
T_2(p) &= -2\sqrt2 {(1-\e)\over  s_{ab}^2}
                   {p\cdot P\,\sand{b}.{\psl}.{a}^2\over s_{{a}\ell}}
             - 2\sqrt2{\sand{b}.{\polsl_P}.{a}\over s_{ab}} 
                   {\sand{b}.{\psl}.{a}\over s_{{a}\ell}}\,.\cr
}\anoneqn$$

\def\AI{J}
The integrals we must perform then include various 
standard tensor two- and three-point integrals, along with integrals of the
following form,
$$\eqalign{
\AI_1(s_{ab},z) &=-i\int {d^{4-2\e} p\over (2\pi)^{4-2\e}}\;
   {1\over p^2 (p-k_a)^2 (p-k_a-k_b)^2 \,p\cdot q}\,,\cr
\AI_2(s_{ab},z) &=-i\int {d^{4-2\e} p\over (2\pi)^{4-2\e}}\;
   {1\over p^2 (p-k_a-k_b)^2 \,p\cdot q}\,,\cr
\AI_3^\mu(s_{ab},z) &=-i\int {d^{4-2\e} p\over (2\pi)^{4-2\e}}\;
   {p^\mu\over p^2 (p-k_a)^2 (p-k_a-k_b)^2 \,p\cdot q}\,,\cr
}\eqn\AxialGaugeLikeIntegrals$$
These integrals resemble those of axial gauge. We emphasize however that
no explicit choice of gauge has been made above, and because all calculations leading
to the cut expression are done on-shell (albeit in $4-2\e$ dimensions), none
is needed.  Furthermore, the splitting amplitudes will be independent of $q$
(as they must be).

The integrals clearly scale like $1/r$ if we scale $q\rightarrow r q$.  
They can
depend only on $q\cdot k_a$ and $q\cdot k_b$ in addition to $s_{ab}$.  Other than
the dot products of $q$ with external momenta, however, $s_{ab}$ is the only
dimensionful parameter.  Furthermore, $q\cdot k_a = z \,q\cdot (k_a+k_b)$ in the
collinear limit, so that the first two integrals necessarily have the form
$$\eqalign{
\AI_1(s_{ab},z) &= {1\over \L -s_{ab}\R^{1+\e}\,q\cdot (a+b)}\,f_1(z)\,,\cr
\AI_2(s_{ab},z) &= {1\over \L -s_{ab}\R^{\e}\,q\cdot (a+b)}\,f_2(z)\,.\cr
}\anoneqn$$

\def\cg{c_\Gamma}
Consider first $\AI_1$.  
To evaluate this integral, introduce Feynman parameters, and perform the
loop integration,
$$\eqalign{
\AI_1(s_{ab},z) &= -12i \int d^4c\;\delta(1-\Sigma_i c_i)\,
   \int {d^{4-2\e} p\over (2\pi)^{4-2\e}}\;
   {1\over \LB -(1-c_4) p^2 +2 p\cdot( c_2 k_a + c_3 k_a + c_3 k_b - c_4 q)
                   - c_3 s_{ab}\RB^4}\,\cr
&= 2{\Gamma(2+\e)\over (4\pi)^{2-\e}} \int d^4c\;\delta(1-\Sigma_i c_i)\,
   {(1-c_4)^{2\e}\over \LB - c_1 c_3 s_{ab}
            -2 c_4 q\cdot(k_a+k_b)\,(c_2 z + c_3)\RB^{2+\e}}\cr
}\anoneqn$$

In order to solve for $f_1(z)$, we may set and $q\cdot(k_a+k_b)/s_{ab}$
to any convenient non-zero value.  It will be convenient to choose 
it to be $1/2$; at this point,
$$\eqalign{
f_1(z) &= -{1\over2}(-s_{ab})^{2+\e} 
   \LP \vphantom{A^A}\AI_1(s_{ab},z)\RV_{2 q\cdot(k_a+k_b)=s_{ab}}
   \cr
 &= -{\Gamma(2+\e)\over (4\pi)^{2-\e}} \int d^4c\;\delta(1-\Sigma_i c_i)\,
   {(1-c_4)^{2\e}\over \LB c_1 c_3 
            + c_2 c_4 z + c_3 c_4\RB^{2+\e}}\,.\cr
}\eqn\fone$$
Using the following change of variables,
$$
c_1 = y (1-x)\,,\hskip 10mm
c_2 = (1-y) w\,,\hskip 10mm
c_3 = (1-y) (1-w)\,,\hskip 10mm
c_4 = y x\,,
\eqn\ChangeJone$$
(with jacobian $y(1-y)$), 
we find
$$\eqalign{
f_1(z) &= -{\Gamma(2+\e)\over (4\pi)^{2-\e}}
      \int_0^1 dx dy dw\;\LB y(1-y)\RB^{-1-\e}
   {(1-x y)^{2\e}\over \LB 1-w+x w z\RB^{2+\e}}\cr
&= {2\over\e^2} \cg \LB -{\Gamma(1-\e)\Gamma(1+\e)}
                            \,z^{-1-\e} (1-z)^\e
                        -{1\over z}
                        +{(1-z)^\e\over z}\,\F21\L \e,\e;1+\e;z\R\RB\cr
&= {2\over\e^2} \cg \LB -{\Gamma(1-\e)\Gamma(1+\e)}
                            \,z^{-1-\e} (1-z)^\e
                        -{1\over z}
                        +{(1-z)^\e\over z}
                         +{\e^2\over z} \Li_2(z) \RB + \Ord(\e)\,,\cr
}\anoneqn$$
where $\F21$ is the Gauss hypergeometric function, 
$\Li_2$ the dilogarithm, and the standard one-loop prefactor is,
$$
\cg = {\Gamma(1+\epsilon)\Gamma^2(1-\epsilon)\over
  (4\pi)^{2-\epsilon}\Gamma(1-2\epsilon)}\,.
\anoneqn$$

The other integral in eqn.~(\use\AxialGaugeLikeIntegrals), 
$\AI_2$, is manifestly $z$-independent, and one finds that
$$
f_2(z) = -{1\over\e^2}\cg\,.
\anoneqn$$
The last integral we need can be obtained by a Passarino--Veltman
reduction~[\use\BrownFeynman,\use\Passarino],
$$
\AI_3^\mu(s_{ab},z) = {1\over 2\,\L -s_{ab}\R^{1+\e}\,q\cdot (a+b)}
\,\LB f_1(z)\,k_a^\mu + {1\over 1-z}\L 2 f_2 - z f_1(z)\R\, k_b^\mu
           +{1\over 1-z}\L f_1(z)-2 f_2\R\, q^\mu
   \RB
\anoneqn$$

With these results, we obtain immediately
$$\eqalign{
 \Cone&(P\rightarrow a^+\,b^-;z) =
\cr &\hskip -8mm
-{1\over 2\sqrt{2}\,s_{ab}} 
   \LB  z f_1(z) + (1-z) f_1(1-z)  -2 f_2\RB\,
\L{\mu^2\over -s_{ab}}\R^{\e}\,
\LB {(1-2 z)\over \sqrt{z (1-z)}} \sand{b}.{\polsl_{P}}.{a}
     + (k_a-k_b)\cdot\pol_P\RB\,.
}\anoneqn$$
The corresponding tree-level result is,
$$
-{1\over \sqrt2 s_{ab}} 
\LB {(1-2 z)\over \sqrt{z (1-z)}} \sand{b}.{\polsl_{P}}.{a}
     + (k_a-k_b)\cdot\pol_P\RB\,.
\anoneqn$$

\section{General Form of Splitting Amplitudes}
\tagsection\GeneralSection
\vskip 10pt

We can also use the basic eqn.~(\use\MasterFormula) for the one-loop
splitting amplitudes in order to obtain a more general expression.
We can repeat the steps in the previous section to obtain
an expression for
the gluon splitting amplitude with arbitrary external polarization
vectors.  In this case, we would use the general four-point amplitude
rather than the expression~(\use\FourPointPM), sum over polarizations
crossing the cut using the identity~(\use\PolarizationSum), and
perform the loop integrals.  Doing so, we find 
(in the CDR scheme) the following result for the one-loop splitting
amplitude,
$$\eqalign{
\Cone&(P\rightarrow a\,b;z) = \cr
&\hskip 10mm
{1\over 2} \L{\mu^2\over -s_{ab}}\R^{\e}
   \LB z f_1(z) + (1-z) f_1(1-z)-2 f_2\RB\,\Ctree(P\rightarrow a\,b;z) \cr
&\hskip 15mm + {1\over\sqrt2 s_{ab}^2}
 {\e^2\over (1-2\e)(3-2\e)}\L{\mu^2\over -s_{ab}}\R^{\e} f_2
         \,(k_a-k_b)\cdot\pol_P\, 
              \L s_{ab} \pol_a\cdot\pol_b - 2 k_b\cdot \pol_a\,k_a\cdot \pol_b\R\,.
}\eqn\GeneralFormA$$
This calculation requires some tensor integrals beyond those presented
in section~\use\ExampleSection; they are listed in appendix~\use\IntegralsAppendix.
As expected, the coefficient proportional to the tree-level structure
has $1/\e^2$ and $1/\e$ poles corresponding to soft and collinear singularities
of the virtual gluons, as well as an ultraviolet singularity (the formula
here is given before ultraviolet subtractions).  The second polarization
structure, which is new at loop level, is in contrast finite when
$\e\rightarrow 0$, as it must be.  The two coefficients of the
two polarization structures will make additional appearances later on,
and so it will be convenient to define
$$\eqalign{
r_1^{[1]}(z) &= {1\over 2} \L{\mu^2\over -s_{ab}}\R^{\e}
   \LB z f_1(z) + (1-z) f_1(1-z)-2 f_2\RB\,,\cr
r_2(z) &=  {\e^2\over (1-2\e)(3-2\e)}\L{\mu^2\over -s_{ab}}\R^{\e} f_2\,.\cr
}\eqn\rGGG$$
The first term in eqn.~(\use\GeneralFormA) is in fact the same in
all variants of dimensional regularization, but the second term is not.
The corresponding results in the four-dimensional helicity scheme (FDH) [\use\FDHS]
may be obtained by correcting for the difference ($2$ versus $2-2\e$)
of helicity states with respect to the CDR scheme.  This difference can
also be expressed in terms of the contribution of an adjoint scalar inside 
the loop; for the general result, we find
$$
r_2^{[1]}(z) =  {(1-\e\delta)
\e^2\over (1-2\e)(1-\e)(3-2\e)}\L{\mu^2\over -s_{ab}}\R^{\e} f_2\,,
\eqn\CDRFDHggg$$
where
$$
\delta = \left\{ \eqalign{&0\quad\hbox{\rm in the FDH scheme,}\cr
  &1 \quad\hbox{\rm in the CDR (or HV) scheme.} }\right.
$$
\def\SecondTensor{{(k_a-k_b)\cdot\pol_P\over\sqrt2 s_{ab}^2}\, 
\L s_{ab} \pol_a\cdot\pol_b 
 - 2 k_b\cdot \pol_a\,k_a\cdot \pol_b\R}
We may write the separate contributions with particles of spin $J$ circulating
using the same polarization tensors,
$$\eqalign{
\delta\Cone_{[J]}&(P\rightarrow a\,b;z) = \cr
&\hskip 2mm
r_1^{[J]}(z) \Ctree(P\rightarrow a\,b;z) 
+r_2^{[J]}(z)\,\SecondTensor\,.
}\eqn\GluonOneLoopPara$$
(The formul\ae\ assume two helicity states, that is Weyl or Majorana fermions
and complex scalars.)

If we now take the external legs to be in four dimensions, we can
write down expressions for the various helicity choices.
For example, the following relations hold,
$$\eqalign{
\Cone&(P^+\rightarrow a^\pm\,b^\mp;z) = \Ctree(P^+\rightarrow a^\pm\,b^\mp;z)
\,r_1(z),\cr
\Cone&(P^-\rightarrow a^+\,b^+;z) = \Ctree(P^-\rightarrow a^+\,b^+;z)
\,(r_1(z)-z(1-z)\,r_2),\cr
\Cone&(P^+\rightarrow a^+\,b^+;z) = 
\sqrt{z(1-z)}{\spb{a}.{b}\over\spa{a}.{b}^2 }\,r_2.
}\eqn\ReltoFDH
$$
The remaining helicity configurations can be obtained using parity.

For completeness, we also record the contribution of a (complex) adjoint scalar 
circulating in the loop,
$$\eqalign{
r_1^{[0]} &=0,\cr
r_2^{[0]} &={\e^2\over (1-2\*\e)(1-\e)(3-2\*\e)}
\L{\mu^2\over -s_{ab}}\R^{\e}\*f_2. 
}
\eqn\scalarggg$$
For the scalars in the fundamental representation of the gauge 
group, these functions must be multiplied by a factor of $1/N$.

\def\qb{{\overline q}}
The basic eqn.~(\use\MasterFormula) also applies to splitting amplitudes
with fermions either circulating in the loop, or as external states.
To compute the contributions to the gluon splitting amplitude proportional
to the number of (massless) flavors $n_f$,
we need the tree-level $g\rightarrow q\qb$ splitting
amplitude (to the left of the cut), and the two-quark two-gluon amplitude
(to the right of the cut),
$$\eqalign{\delta \Cone_{[1/2]}(P\rightarrow a\,b;z) &=
\sum_{\phpol} 
\int {d^{4-2\e}\ell\over (2\pi)^{4-2\e}}\;{i\over\ell^2}
  \Ctree(P\rightarrow (\ell+a+b)_f\,(-\ell)_f){i\over (\ell+k_a+k_b)^2}
\cr &\hskip 40mm\times
      A_{4}^\tree((-\ell-a-b)_f,a,b,\ell_f)
}\anoneqn$$
where the $f$ subscript indicates fermionic legs.  
Evaluating this expression, we obtain
for the coefficient of the two tensor structures,
$$\eqalign{
r_1^{[1/2]} &=0,\cr
r_2^{[1/2]} &=
-{\e^2\over (1-2\*\e)(1-\e)(3-2\*\e)}
\L{\mu^2\over -s_{ab}}\R^{\e}\*f_2.
}
\eqn\fermionggg$$
The contributions proportional to the number of flavors is thus
$$
\LP\delta \Cone(P\rightarrow a\,b;z)\RV_{n_f} =
{n_f\over N}\SecondTensor\, r_2^{[1/2]}\,.
\anoneqn$$
The $1/N$ factor
reflects the quarks being in the fundamental rather than the
adjoint representation.  

\def\lc{{\rm LC}}
\def\slc{{\rm SC}}
The one-loop $g\rightarrow q\qb$ splitting amplitude requires, in
addition to these tree-level splitting and four-point amplitudes,
the four-quark amplitude.  Here, there are two contributions which
are independent of the number of flavors.   One comes from gluons crossing
the cut; the other, subleading in the number of colors, from 
quarks crossing the cut.  The former is,
$$\eqalign{
&\LP\delta \Cone(P\rightarrow a_q\,b_\qb;z)\RV_{\lc} =
\cr &\hskip 20mm
\sum_{\phpol\ \ll1,\ll2} 
\int {d^{4-2\e}\ell\over (2\pi)^{4-2\e}}\;{i\over\ell^2}
  \Ctree(P\rightarrow (\ell+a+b)\,(-\ell)){i\over (\ell+k_a+k_b)^2}
\cr &\hskip 50mm\times \vphantom{\sum_L}
      A_{4}^\tree((-\ell-a-b),a_q,b_\qb,\ell)\,.
}\eqn\gqqgluonsMaster$$

\def\ub{{\overline u}}
The contribution from quarks crossing the cut is given by the following
equation,
$$\eqalign{
&\LP\delta\Cone(P\rightarrow a_q\,b_\qb;z)\RV_{\slc} =
\cr &\hskip 20mm
  {1\over N^2}\sum_{\phpol\ \ll1,\ll2} 
\int {d^{4-2\e}\ell\over (2\pi)^{4-2\e}}\;{i\over\ell^2}
  \Ctree(P\rightarrow (\ell+a+b)_q\,(-\ell)_{\qb}){i\over (\ell+k_a+k_b)^2}
\cr &\hskip 40mm\times\vphantom{\sum_L}
      A_{4}^\tree((-\ell-a-b)_{\qb'},a_{q'},b_\qb,\ell_q)
}\anoneqn$$
where the subscript $q$ indicates a quark,
and $\qb$, an antiquark.  
Although the external quark and antiquark must
necessarily have the same flavor, we make use
of four-quark amplitudes with two different flavors in order
to isolate the $n_f$-independent contribution.  We find for the two 
different contributions,
$$\eqalign{
  \LP\delta \Cone(P\rightarrow a_q\,b_\qb;z)\RV_{\lc} &=\cr
  &\hskip-3.5cm\Ctree(P\rightarrow a_q\,b_\qb;z)
 \bigg\{r_1^{[1]}(z)-\bigg[{1\over 1-2\*\e}
  -{(1-\delta\*\e)\*\e\over 2(1-\e)}
  +{2\*\e\*(1-\e)\*(1-\delta\*\e)\over(1-2\*\e)\*(3-2\*\e)}
  \bigg]\*\L{\mu^2\over -s_{ab}}\R^{\e}\*f_2\bigg\}
\cr
\LP\delta\Cone(P\rightarrow a_q\,b_\qb;z)\RV_{\slc} &=
-{1\over N^2}\Ctree(P\rightarrow a_q\,b_\qb;z)
 \L{1\over 1-2\*\e}-{(1-\delta\*\e)\*\e\over 2\*(1-\e)}\R
  \L{\mu^2\over -s_{ab}}\R^{\e}\*f_2
}\eqn\gqqgluons$$
This splitting amplitude is again scheme-dependent; to compute the
difference between the CDR and FDH schemes one can again either
modify the gluon sums over polarizations or equivalently compute
the contributions of $(-\e)$ complex internal scalars.  In the latter
method, the Yukawa coupling must be adjusted to obtain the `same'
coupling as for a gluon.  The adjustment gives precisely the Yukawa
coupling that would be obtained in an $N=4$ supersymmetric gauge
theory.  The relevant scalar contribution, combining leading-
and subleading-color parts, is
$$\eqalign{
  \LP\delta \Cone_{[0]}(P\rightarrow a_q\,b_\qb;z)\RP &=\cr
  &\hskip-4cm\Ctree(P\rightarrow a_q\,b_\qb;z)
  \L{\e\over (1-2\*\e)\*(3-2\*\e)}
  -{\e\over 2\*(1-\e)\*(1-2\*\e)}
  +{1\over N^2}{\e\over 2(1-\e)}\R\L{\mu^2\over -s_{ab}}\R^{\e}\*f_2
\cr  &\hskip-4cm=\Ctree(P\rightarrow a_q\,b_\qb;z)
  \L
  {\e\over2(1-\e)}-{2\*\e\*(1-\e)\over (1-2\*\e)\*(3-2\*\e)}
  +{1\over N^2}{\e\over 2(1-\e)}\R\L{\mu^2\over -s_{ab}}\R^{\e}\*f_2\,.
}\eqn\gqqscalars$$
In the first form, the last two terms arise from the Yukawa coupling.
The second form matches the scheme-dependent parts shown in
eqn.~(\use\gqqgluons).  In the above equations, the tree splitting
amplitude is
$$
\Ctree(P\rightarrow a_q\,b_\qb;z) = {1\over\sqrt2 s_{ab}} \ub_a \polsl_P u_b\,.
\anoneqn$$
Unlike the pure gluon case, for this splitting amplitude no
other structure is possible when the fermions are kept on-shell.

The contribution to the $g\rightarrow q\qb$ splitting
amplitude proportional to the number of flavors is,
$$\eqalign{
\LP\delta\Cone(P\rightarrow a_q\,b_\qb;z)\RV_{n_f} &=
{n_f\over N}
\cr &\hskip -30mm \times \hskip -5mm\sum_{\phpol\ \ll1,\ll2} 
\int {d^{4-2\e}\ell\over (2\pi)^{4-2\e}}\;{i\over\ell^2}
  \Ctree(P\rightarrow (\ell+a+b)_f\,(-\ell)_f){i\over (\ell+k_a+k_b)^2}
\cr &\hskip 40mm\times\vphantom{\sum_L}
      A_{4}^\tree((-\ell-a-b)_f,a_q,b_\qb,\ell_f)
\cr
&= {n_f\over N}
\Ctree(P\rightarrow a_q\,b_\qb;z)
{2\*\epsilon\*(1-\epsilon)\over(1-2\*\epsilon)(3-2\*\epsilon)}
 \L{\mu^2\over -s_{ab}}\R^{\e} f_2\,.
}\eqn\gqqfermlc$$

We turn next to the fermion splitting amplitude, $q\rightarrow qg$.
All contributions here involve a tree-level $q\rightarrow qg$
splitting amplitude to the left of the cut, and a two-quark two-gluon
amplitude to the right; but there are both leading- and subleading-color
ones.  The leading-color term is,
$$\eqalign{
\Cone(P_q\rightarrow a_q\,b;z) &=
\sum_{\phpol\ \ll1,\ll2} 
\int {d^{4-2\e}\ell\over (2\pi)^{4-2\e}}\;{i\over\ell^2}
  \Ctree(P_q\rightarrow (\ell+a+b)_f\,(-\ell)){i\over (\ell+k_a+k_b)^2}
\cr &\hskip 40mm\times
      A_{4}^\tree((-\ell-a-b)_f,a_q,b,\ell)\,.
}\eqn\Masterqg$$ 
The result can be written in terms of the tree-level splitting amplitude,
along with a new polarization tensor,
\def\ump{u_{-\!{P}}}
$$\eqalign{
\Ctree(P_q\rightarrow a_q\,b;z) &= 
-{1\over\sqrt2 s_{ab}} \ub_a \polsl_P \ump \,,\cr
  \Cone(P_q\rightarrow a_q\,b;z) &=
  \Ctree(P_q\rightarrow a_q\,b;z) r_3^{\lc} 
  - {\sqrt{2}\over s^2_{ab}} \ub_a \ksl_b \ump
  \, k_a\cdot\pol_b r_4^{\lc}\,.
}\eqn\qgstructure$$
Note that the momentum $P$ is outgoing (in accord with our overall 
convention), and hence $\ump$ is the usual spinor for an inward-directed
momentum.  Given our conventions, helicity conservation means that the
two external fermions at the ends of a given fermion line have {\it opposite\/}
helicities: flipping the direction of a momentum also flips the sign
of the helicity.  

The functions $r_{3,4}$ appearing in eqn.~(\use\qgstructure) are,
$$\eqalign{
r_3^{\lc} &= {1\over 2}
  \L(1-z)f_1(1-z)-{(1-\delta\*\e)\e^2\over (1-\e)(1-2\*\e)}\*f_2
  \R\L{\mu^2\over -s_{ab}}\R^{\e}\,,\cr
r_4^{\lc} &= {1\over 2}
  {(1-\delta \e)\e^2\over (1-\e)(1-2\e)}\L{\mu^2\over -s_{ab}}\R^{\e}f_2 \,.
}\eqn\rthreefourlc$$

The subleading-color terms are given by a formula similar to~(\use\Masterqg),
but with the fermion legs not adjacent in the four-point amplitude to the
right of the cut,
$$\eqalign{
\LP\delta \Cone(P_q\rightarrow a_q\,b;z)\RV_{\slc} &=
\cr &\hskip -30mm
-{1\over N^2} \sum_{\phpol\ \ll1,\ll2} 
\int {d^{4-2\e}\ell\over (2\pi)^{4-2\e}}\;{i\over\ell^2}
  \Ctree(P_q\rightarrow (\ell+a+b)_f\,(-\ell)){i\over (\ell+k_a+k_b)^2}
\cr &\hskip 40mm\times
      A_{4}^\tree((-\ell-a-b)_f,a_q,\ell,b).
}\anoneqn$$

We may express the result of the subleading-color contributions
in a decomposition similar to that in eqn.~(\use\qgstructure),
but multiplied by an overall factor of $1/N^2$.
The coefficient functions are,
$$\eqalign{
r_3^{\slc} &= -{1\over 2}
  \bigg(
  z f_1(z)-2\*f_2+{(1-\delta\*\e)\*\e^2\over (1-\e)\*(1-2\*\e)}
  \*f_2
  \bigg)\L{\mu^2\over -s_{ab}}\R^{\e},\cr
r_4^{\slc} &= {1\over 2}
  {(1-\delta\*\epsilon)\*\epsilon^2\over (1-\epsilon)(1-2\*\epsilon)}
  \L{\mu^2\over -s_{ab}}\R^{\e}\*f_2.
}\eqn\rthreefoursc$$

The difference between schemes again turns out to be expressible using contributions
in which the gluon crossing the cut is replaced by a scalar; then
for both the leading-color and subleading-color terms, these
contributions are,
$$\eqalign{
r_3^{[0]} &= -{\e^2\over 2(1-\e)\*(1-2\*\e)}
\*\L{\mu^2\over -s_{ab}}\R^{\e}\*f_2\,,\cr
r_4^{[0]} &= {\e^2\over 2(1-\e)\*(1-2\*\e)}
\*\L{\mu^2\over -s_{ab}}\R^{\e}\*f_2 \,.
}\anoneqn$$
(The subleading-color contributions are of course again multiplied by
a factor of $1/N^2$, not included here.)

For specific helicity choices for the external legs, we then find,
$$\eqalign{
\Cone&(P^-_q\rightarrow a^+_q\,b^-;z) =\Ctree(P^-_q\rightarrow a^+_q\,b^-;z)
(r_3+r_4)\,,\cr 
\Cone&(P^-_q\rightarrow a^+_q\,b^+;z) =\Ctree(P^-_q\rightarrow a^+_q\,b^+;z)
  (r_3+z\,r_4)\,.\cr
}\eqn\ReltoFDHqg$$

\section{Consistency Checks}
\tagsection\CheckSection
\vskip 10pt

The splitting amplitudes computed in the previous section may appear
to be completely independent quantities.  They are not.  If
we set $n_f/N\rightarrow1$ and $1/N^2\rightarrow -1$, we obtain
the corresponding quantities in a supersymmetric theory.  These
latter satisfy various identities descended from the supersymmetry
Ward identity [\use\SWI].  Such identities thus provide a cross-check
on the splitting amplitudes even in nonsupersymmetric theories
such as QCD.

\def\gt{{\tilde g}}
Supersymmetry Ward identities have long been recognized [\use\UseSWI,\use\ManganoReview]
as providing useful relations between different amplitudes in
gauge theories.  For example, they relate amplitudes with
$n$ gluons to those with $(n-2)$ gluons and two gluinos.
For certain choices of the external helicities, these
identities, which hold to all orders in perturbation
theory, allow us to solve for the two-gluino amplitudes
in terms of the $n$-gluon ones.   For example,
$$\eqalign{
A^\tree_n&(1_\gt^+,2^-,3^+,\ldots,j_\gt^-,\ldots,n^+)
 =  
-{\spa1.2\over\spa2.j}
A^\tree_n(1^+,2^-,3^+,\ldots j^-,\ldots,n^+)\,.
}\eqn\SWIexample$$
At tree level, the two-gluino color-ordered amplitudes are in fact
identical to two-quark ones, so the identity links two different
QCD color-ordered amplitudes.  Beyond tree level, color-ordered QCD amplitudes
are of course no longer identical to those in a supersymmetric theory.
Supersymmetry identities are nonetheless still useful in relating different
contributions to QCD amplitudes [\use\AnnRev].  

The splitting amplitudes are in fact identical for all matterless supersymmetric
gauge theories, independent of the number of supersymmetries.  We may therefore
pick an $N=1$ supersymmetric gauge theory, with no matter content, to 
study the supersymmetry identities.

What relations do we expect for the splitting amplitudes?
If we consider the $n=5$ case of eqn.~(\use\SWIexample), with legs
relabeled, we obtain the following relation between an all-gluon
and a two-gluino three-gluon amplitude,
$$
  \spa3.1 A_5(1^+,2^+,3^-,4^+,5^-)
  = \spa3.5 A_5(1_\gt^+,2^+,3^-,4^+,5_\gt^-)\,.
\eqn\SUSYid
$$
This relation holds to all orders in perturbation theory.
Let us now examine the collinear limit $k_1\parallel k_5$.
At tree-level, we then obtain the identity ($P =-k_1-k_5$)
$$\eqalign{
  \spa3.1  A^\tree(2^+,3^-,4^+,(-P)^-) 
  &\Ctree(P^{+}\rightarrow 5^-\,1^+;z)
  = \cr
  &\spa3.5 
  A^\tree(2^+,3^-,4^+,(-P)^{-})
 \Ctree(P^{+}\rightarrow 5_\gt^-\,1_\gt^+;z)\,.
}\anoneqn$$
(Note that the pure gluon amplitudes with a lone negative helicity vanish
to all orders in the supersymmetric theory. The sum over the polarizations 
of the fused leg $k_1+k_5$ thus reduces to a single term.)  This identity
in turn implies a relation between the $g\rightarrow g g$ and
$g\rightarrow \gt\gt$ splitting amplitudes,
$$
\Ctree(P^{+}\rightarrow a_\gt^-\,b_\gt^+;z)
=
\sqrt{1-z\over z}
\Ctree(P^{+}\rightarrow a^-\,b^+;z),
\eqn\gluontogluinogluinosusyidtree$$
where we have used $k_5=z(k_1+k_5)$ and $k_1 = (1-z)(k_1+k_5)$ in the collinear
limit to evaluate the ratio $\spa3.1/\spa3.5$.

If we now repeat this exercise for the one-loop amplitudes, again taking
the limit of eqn.~(\use\SUSYid), we obtain an identity of the same form for the loop
splitting amplitudes,
$$
\Cone(P^{+}\rightarrow a_\gt^-\,b_\gt^+;z)
=
\sqrt{1-z\over z}
\Cone(P^{+}\rightarrow a^-\,b^+;z)\,.
\eqn\gluontogluinogluinosusyidloop$$
The similarity is quite natural as the Ward identities hold to all
orders in perturbation theory.

One subtlety that arises in adapting the computations of the previous
section to use in a supersymmetric theory is a difference in phase conventions.
It is therefore more convenient to take the ratio of the tree-level
and one-loop identities,
$$
{\Cone(P^{+}\rightarrow a_\gt^-\,b_\gt^+;z)\over 
\Ctree(P^{+}\rightarrow a_\gt^-\,b_\gt^+;z)} =
{\Cone(P^{+}\rightarrow a^-\,b^+;z)\over 
  \Ctree(P^{+}\rightarrow a^-\,b^+;z)}\,,
\eqn\firstonetocheck$$
as a test of our computation.   This form of the identity is not dependent
on the relative phase conventions for gluons and gluinos.

In performing this check, we should of course use a supersymmetry-preserving
regulator, so we pick the FDH scheme ($\delta=0$).  

To obtain the $g\to \tilde g\tilde g$ splitting amplitude, start
with the $g\to q\qb$ amplitude~(\use\gqqgluons,\use\gqqfermlc),
$$\eqalign{
\Ctree(P\rightarrow a_q\,b_\qb;z)
  \,
  \bigg[r_1^{[1]}(z)
  &-2{\*(1-\e)\*(1-n_f/N-\delta\*\e)\*\e\over (1-2\*\e)\*(3-2\*\e) }
  \L{\mu^2\over -s_{ab}}\R^{\e}\*f_2
\cr &-\Big(1+{1\over N^2}\Big)\,\Big[ {1\over1-2\e} - {(1-\delta\e)\over 2(1-\e)}\Big]
  \L{\mu^2\over -s_{ab}}\R^{\e}\*f_2
  \bigg]\,,
}\anoneqn$$
and set $n_f/N\rightarrow 1$,
$1/N^2\rightarrow -1$, and $\delta=0$:
$$
\Cone(P\rightarrow a_\gt\,b_\gt;z) =
\Ctree(P\rightarrow a_\gt\,b_\gt;z)
  \, r_1^{[1]}(z).
\eqn\gtogluinogluino$$

In an $N=1$ supersymmetric theory, the fermionic contribution to
$g\to gg$ cancels the $r_2^{[1]}$ term in the gluon contribution, and so we obtain,
$$
\Cone(P\rightarrow a\,b;z) = \Ctree(P\rightarrow a\,b;z)\, r_1^{[1]}(z)\,.
\eqn\gtoggSUSY$$
Both sides of eqn.~(\use\firstonetocheck) are thus equal to $r_1^{[1]}(z)$, 
and the identity is obeyed.

The vanishing of $r_2^{[1]}+r_2^{[1/2]}$ is in fact also a result of
a supersymmetry Ward identity.  To see this, make use of the following
Parke--Taylor equation~[\use\ParkeTaylor],
$$\eqalign{
A_n(1^+,2^+,\ldots,n^-) &= 0\,,
}\anoneqn$$
which are consequences of the Ward identities along with helicity
conservation for fermionic amplitudes.  Consider the collinear limit 
$k_1\parallel k_2$ of this equation at tree level.  Using the above equation itself,
the polarization sum reduces to a lone term ($P=-(k_1+k_2)$),
$$
A_n^\tree((-P)^-,3^+,\ldots,n^-) \Split(P^+ \to 1^+\,2^+;z) = 0\,.
\anoneqn$$
Since the amplitude does not vanish, the splitting amplitude must.
Repeating the above limit for loop amplitudes,
we see that this result holds to all orders in perturbation theory.
But in an $N=1$ supersymmetric theory, this implies that
$$
\LP\Split(P^+\rightarrow a^+\,b^+;z)\RV_{[1]} 
+ \LP\Split(P^+\rightarrow a^+\,b^+;z)\RV_{[1/2]} = 0\quad{\rm or}\quad
 r_2^{[1]}=-r_2^{[1/2]}\,.
\anoneqn$$
If we consider theories with additional supersymmetries, or massless
matter multiplets in an $N=1$ theory, we can also relate the contribution
of internal scalars to the fermionic contribution:
$$
\LP\Split(P^+\rightarrow a^+\,b^+;z)\RV_{[1/2]} 
+ \LP\Split(P^+\rightarrow a^+\,b^+;z)\RV_{[0]} = 0\quad{\rm or}\quad
 r_2^{[1/2]}=-r_2^{[0]}.
\anoneqn$$
Inspection of eqns.~(\use\scalarggg) and~(\use\fermionggg) shows
that the results presented herein indeed satisfy this relation.

We can also derive a relation between 
the $g\to gg $ and the $\tilde g\to\tilde g g$ splitting amplitudes. 
To do so consider the collinear limit $k_4||k_5$ 
of eqn.~(\use\SUSYid) at tree-level ($P=-k_4-k_5$),
$$\eqalign{
  \spa3.1 A^\tree(1^+,2^+,3^-,(-P)^-)
  &\Ctree(P^+\rightarrow 4^+\,5^- ;z )
  = \cr
  &\spa3.5 A^\tree(1^+_{\gt},2^+,3^-,(-P)^-_\gt)
  \Ctree(P^+_\gt\rightarrow 4^+\,5^-_\gt ;z)\,.
}\anoneqn$$
Using the relation
$$
  \spa3.1 A(1^+,2^+,3^-,4^-)
  = \spa3.4 A(1^+_\gt,2^+,3^-,4^-_\gt)\,,
\anoneqn$$
we obtain
$$
  \Split(P^+_\gt\rightarrow a^+\,b^-_\gt ;z)
  ={1\over \sqrt{1-z}} \Split(P^+\rightarrow a^+\,b^- ;z )
  .
\eqn\gluinotogluinogluonid
$$
We have omitted the superscript `tree' because repeating the limit
for loop amplitudes we find that it holds for them as well.
To avoid problems with phase conventions, it is again better to form
ratios,
$$
{\Cone(P^+_\gt\rightarrow a^+\,b^-_\gt ;z)
\over \Ctree(P^+_\gt\rightarrow a^+\,b^-_\gt ;z)}
=
{\Cone(P^+\rightarrow a^+\,b^- ;z )\over\Ctree(P^+\rightarrow a^+\,b^- ;z )}\;.
\eqn\secondonetocheck
$$

Setting $1/N^2\rightarrow -1$ in
eqns.~(\use\qgstructure,\use\rthreefourlc,\use\rthreefoursc)
we obtain,
$$
\Cone(P_\gt\rightarrow a_\gt\,b;z) =
  \Ctree(P_\gt\rightarrow a_\gt\,b;z)\, r_1^{[1]}(z)
\eqn\gluinotogluinogluon
$$
for the $\gt \to \gt g$ splitting amplitude, with $r_1^{[1]}(z)$ defined
in eqn.~(\use\rGGG). 
As a result of setting $1/N^2\to -1$, the function multiplying the
structure
$$\ub_a \ksl_b \ump
  \, k_a\cdot\pol_b
\anoneqn$$
(new at the one-loop level) vanishes. 
We are thus left with only one function needed to describe the one loop result 
for the $\tilde g \to \tilde g g$  splitting amplitude, independent of
the external helicities.
We note that although one should use a supersymmetry-preserving regularisation scheme
(i.e. FDH) it is remarkable that the difference between the results in 
the FDH and CDR schemes
cancels out in the expression above. 

Comparing with eqn.~(\use\gtoggSUSY), we see that this second supersymmetry
identity is also satisfied.

Given the observation that in supersymmetric theories, only one universal function
multiplying the tree-level results appears in the one-loop splitting
amplitudes, the two checks described 
above for a specific combination of 
helicities are sufficient to check the supersymmetry identities for all possible
helicities. 

We may also compare our results with results
previously obtained in the literature.
In particular,
we will compare with the result of 
Bern, Del Duca, and Schmidt [\use\BDS] for the $g\to gg$ splitting 
amplitude to all orders in $\e$, 
and the results (to $\Ord(\e^0)$) for the $g\to q\qb$ and $q\to qg$ 
splitting amplitudes quoted in ref.~[\use\SusyFour]. 
For the comparison with ref.~[\use\BDS] it is convenient 
to use a slightly different representation for
$f_1(z)$ than the one given in eqn.~(\use\fone). 
Transforming the arguments
of the hypergeometric function by the use of two identities, 
eqns.~(9.131.1) and~(9.132.2) of ref.~[\use\GR] we may write 
$$
zf_1(z) =  {2\over\e^2}\cg
\LB
-\Gamma(1-\e)\Gamma(1+\e)\L1-z\over z\R^\e
-{z\over 1-z}{\e\over 1+\e}
\F21(1,1+\e;2+\e;{z \over z-1})
\RB,
\anoneqn$$
and
$$
(1-z)f_1(1-z) 
=  -{2\over\e^2}\cg
\F21(1,-\e;1-\e;{z\over z-1  }).
\anoneqn$$
We may now expand the hypergeometric functions in terms of
polylogarithms, to obtain
$$
-{z\over 1-z}{\e\over 1+\e}
\F21(1,1+\e;2+\e;{z \over z-1})-\F21(1,-\e;1-\e;{z\over z-1  })
= 2\sum_{k=1,3,5,\ldots}^\infty\e^{k} \Li_k\L{-z\over 1-z}\R-1\,,
\anoneqn$$
and thence,
$$
r_1^{[1]}(z)=\cg{1\over\e^2}
\L{\mu^2\over -s_{ab}}\R^{\e}
\bigg[
-\Gamma(1-\e)\Gamma(1+\e)\,\L1-z\over z\R^\e
+2\sum_{k=1,3,5,\ldots}^\infty \e^{k} \Li_k\L{-z\over 1-z}\R
\bigg]\,,
\eqn\roneBernetal$$
where the polylogarithms are defined as follows [\use\Lewin],
$$\eqalign{
\Li_1(x) &= -\ln(1-x)\,,\cr
\Li_k(x) &=\int\limits_0^x {dt\over t}\Li_{k-1}(t) \quad(k=2,3,\ldots)\,.
}\anoneqn$$
Comparing eqn.~(5) of ref.~[\BDS] with eqn.~(\use\ReltoFDH),
we see that comparing our results with theirs reduces to checking
the following pair of equations,
$$\eqalign{
G^n &=r_1^{[1]}(z),\cr
G^f &=-z(1-z)\L r_2^{[1/2]}+r_2^{[1]}\R\,.
}\eqn\beetalvskoetal$$
Comparing eqn.~(\use\roneBernetal), and 
eqns.~(\use\CDRFDHggg, \use\fermionggg) with the results presented in 
ref.~[\use\BDS], it is easy to see that the one-loop
results shown in this paper indeed satisfy eqn.~(\use\beetalvskoetal)
and thus agree
with the result\footnote{${}^{\dagger}$}{
Note that there is misprint in eqn.~(9) of the journal version of
ref.~[\use\BDS] due to 
a misplaced line break. One should replace 
$\Gamma\times (1+\e)$ by $\Gamma(1+\e)$.}
 given by Bern, Del Duca and Schmidt to all orders in 
the dimensional regularization parameter $\e$. 
The contribution of an internal scalar is related to the
difference between the FDH and CDR schemes
in the gluon splitting amplitudes.  Since our results agree
with ref.~[\BDS] for both schemes, we implicitly agree on
the scalar contribution as well.

The representation of $r_1^{[1]}$ given in eqn.~(\use\roneBernetal)
is also useful for comparing the results
for the $g\to q\qb$ splitting amplitudes with the results
quoted in the literature [\use\SusyFour]. Using 
$$
\Gamma(1-\e)\Gamma(1+\e) = 1 + {\pi^2\over 6}\e^2 + O(\e^4)\,,
\anoneqn$$
we obtain from eqn.~(\use\roneBernetal)
$$\eqalign{
r_1^{[1]}(z) &={\cg\over\e^2}
\L{\mu^2\over -s_{ab}}\R^{\e}
\bigg[ -1+\e\*\ln(1-z)+\e\*\ln(z) - {\e^2\over 2}\*\ln^2({1-z\over z})
- {\pi^2\over 6}\e^2
+O(\e^3)
\bigg]\cr
&={\cg\over\e^2}
\L{\mu^2\over -s_{ab}}\R^{\e}
\bigg[ 1-\L{1\over z}\R^\e-\L{1\over 1-z}\R^\e + \e^2\*\ln(z)\*\ln(1-z)
- {\pi^2\over 6}\e^2
+O(\e^3)
\bigg]\,,
}\anoneqn$$
for the expansion in $\e$.
Together with the expansion of the prefactors,
$$\eqalign{
2\*\epsilon\*(1-\epsilon)(1-\delta\e)\over(1-2\*\epsilon)(3-2\*\epsilon)
&={2\over 3}\*\e+\L{10\over 9}-{2\over 3}\*\delta\R\*\e^2+O(\e^3)\,,\cr
{1\over 1-2\*\e}-{(1-\delta\*\e)\*\e\over 2\*(1-\e)} &=
1+{3\over 2}\*\e+\L{7\over 2}+{1\over2}\*\delta\R\*\e^2+O(\e^3)\,,\cr
{\e\over (1-2\*\e)\*(3-2\*\e)}
&= {1\over 3}\*\e+{8\over 9}\*\e^2+O(\e^3)\,,\cr
}\anoneqn$$
appearing in eqn.~(\use\gqqgluons), (\use\gqqfermlc)
and (\use\gqqscalars), this leads immediately to the 
following expansions 
in $\e$ for the $g\to q\qb$ splitting amplitudes,
$$\eqalign{
\LP\delta\Cone(P\rightarrow a_q\,b_\qb;z)\RV_{n_f} &=
{n_f\over N}\cg
\Ctree(P\rightarrow a_q\,b_\qb;z)
\LB-{2\over 3}{1\over\e}\L{\mu^2\over -s_{ab}}\R^{\e}-{10\over 9}-O(\e)\RB
\cr
\LP\delta\Cone(P\rightarrow a_q\,b_\qb;z)\RV_{\slc} &=
-{1\over N^2}\*\cg\Ctree(P\rightarrow a_q\,b_\qb;z)\cr
 &\LB
 -{1\over\e^2}\L{\mu^2\over -s_{ab}}\R^{\e}
 -{3\over 2}{1\over\e}\L{\mu^2\over -s_{ab}}\R^{\e}
 -\L{7\over 2}+{1\over2}\*\delta\R+O(\e)
 \RB
  \cr
\LP\delta \Cone(P\rightarrow a_q\,b_\qb;z)\RV_{\lc} &=
\cg\Ctree(P\rightarrow a_q\,b_\qb;z)\cr
  &\hskip-3.5cm 
\bigg\{
-{1\over\e^2}\bigg[ \L{\mu^2\over z(-s_{ab})}\R^{\e}
+\L{\mu^2\over (1-z)(-s_{ab})}\R^{\e} 
-2\L{\mu^2\over -s_{ab}}\R^{\e}\bigg] \cr
&+{13\over 6}\*{1\over\e}\L{\mu^2\over -s_{ab}}\R^{\e} 
+ \ln(z)\*\ln(1-z)
- {\pi^2\over 6}
+{83\over 18}-{1\over 6}\*\delta+O(\e)
 \bigg\}
}\anoneqn$$
For the contribution from virtual scalars we obtain,
$$\eqalign{
  \LP\delta \Cone_{[0]}(P\rightarrow a_q\,b_\qb;z)\RP &=
  \cg\Ctree(P\rightarrow a_q\,b_\qb;z)
  \LB
  -{1\over 3}{1\over \e}\L{\mu^2\over -s_{ab}}\R^{\e}
  -{8\over 9}+O(\e)
  \RB\,,
}\anoneqn$$
if we consider only the first term in eqn.~(\use\gqqscalars). 
(Recall
that the two other terms in eqn.~(\use\gqqscalars) are due to the
coupling of the scalars to quarks.) 
These expansions in $\e$ agree with the results 
for the 
$g\to q\qb$ splitting amplitudes 
presented in eqn.~(B.12) of ref.~[\use\SusyFour].

We close these comparisons with that for the $q\to qg$ splitting amplitude.
For this purpose, we must expand the functions
$r_3^{\lc,\slc}$, $r_4^{\lc,\slc}$ in $\e$. This can be easily done using the
following expression for the function $zf_1(z)$,
$$\eqalign{
z f_1(z) &= {2\over\e^2} \cg 
\LB 
-\L{1\over z}\R^\e 
+ \e^2\L\Li_2(z) + \ln(z)\ln(1-z) - {\pi^2\over 6}\R
\RB + \Ord(\e)\cr
&= {2\over\e^2} \cg 
\LB 
-\L{1\over z}\R^\e 
-\e^2\Li_2(1-z)
\RB + \Ord(\e)\,,\cr
}
\anoneqn$$
where we have used the dilogarithm identity
$$
\Li_2(x)+\Li_2(1-x)= {\pi^2\over 6}-\ln(x)\ln(1-x)\,.
\anoneqn$$
The connection between the functions $r_3^{\lc,\slc}$, $r_4^{\lc,\slc}$ used
in this paper and the functions $r_{\rm S}(q^+,a^-)$, $r_{\rm S}(q^+,a^+)$ 
introduced in eqn.~(B.7) of ref.~[\use\SusyFour]
can be read off from eqn.~(\use\ReltoFDHqg):
$$\eqalign{
\cg r_{\rm S}(q^+,a^-) &= r_3^{\lc}+ r_4^{\lc}
      +{1\over N^2} \L r_3^{\slc}+ r_4^{\slc}\R\cr
\cg r_{\rm S}(q^+,a^+) &= r_3^{\lc}+ r_4^{\lc}
      +{1\over N^2} \L r_3^{\slc}+ r_4^{\slc}\R
-(1-z)\,\Bigl(r_4^{\lc}+{r_4^{\slc}\over N^2}\Bigr)
}\eqn\BeetalvsKoetalqqg$$
Using the expansion for $zf_1(z)$ we obtain
$$\eqalign{
r_3^{\lc}+ r_4^{\lc}+{1\over N^2}\L r_3^{\slc}+ r_4^{\slc}\R &=
   \cg 
  \LB 
  -{1\over\e^2}\L{\mu^2\over (1-z)(-s_{ab})}\R^\e 
  -\Li_2(z)
  \RB\cr
  &\hskip 5mm-  \cg{1\over N^2}
  \LB 
  -{1\over\e^2}\L{\mu^2\over z(-s_{ab})}\R^\e 
  +{1\over \e^2}\L{\mu^2\over -s_{ab}}\R^{\e}
  -\Li_2(1-z)
  \RB + \Ord(\e)\,,\cr
-(1-z)\,\Bigl(r_4^\lc+{1\over N^2} r_4^\slc\Bigr) &=
\cg(1+{1\over N^2}){1\over 2}(1-z) + \Ord(\e)\,.
}\anoneqn$$
These results agree with eqns.~(B.10) and~(B.11) of ref.~[\use\SusyFour].

\section{The One-Loop Antenna Function}
\tagsection\AntennaSection
\vskip 10pt

Gauge-theory amplitudes are also singular in the soft limit, 
when a gluon four-momentum vanishes.  Color-ordered amplitudes display
a factorization similar to that in eqn.~(\use\OneLoopCollinearFactorization),
$$\eqalign{
 A_n^{\oneloop}&(1,\ldots,a,s^\ls,b,\ldots,n) 
\;{\buildrel k_s\rightarrow 0\over{\relbar\mskip-1mu\joinrel\longrightarrow}}\cr
& \Soft^\tree(a,s^\ls,b)\,
      A_{n-1}^\oneloop(1,\ldots,a,b,\ldots,n)
\cr &\hskip 20mm  
  +\Soft^\oneloop(a,s^\ls,b)\,
      A_{n-1}^\tree(1,\ldots,a,b,\ldots,n) \;,
}\eqn\OneLoopSoftFactorization$$
in which the collinear splitting amplitude $\Split$ is replaced by the
eikonal amplitude $\Soft$.  The eikonal limit also demonstrates one
of the advantages of the looking at color-ordered amplitudes, namely
that they allow a simple and straightforward factorization.  The amplitude
as a whole does not factorize simply, because color-charge factors get tangled
up with the eikonal function.  The eikonal function may be extracted
to $\Ord(\e)$ from known results for five-gluon 
amplitudes.  It has also been computed to all orders in $\e$ by
Bern, Del~Duca, and Schmidt~[\use\BDS].

In some applications, for example the construction of jet programs, one
wishes to integrate over the singular regions.  One can do this by
integrating over the collinear and soft regions, and using in each
the factorization appropriate to it.  However, one must then introduce
an artificial boundary between the two.  It would be nicer to have
a single, unifying function which summarizes the behavior in all
singular regions.  

Catani and Seymour~[\use\CS] wrote down just such a function,
which appears in a {\it dipole\/} factorization
formula.  They presented a single function capturing the singular behavior
of the squared matrix element in both the soft and collinear limits.
One of us then gave a derivation of a single {\it antenna\/} factorization
function~[\use\Antenna] --- so-called because it is within each color antenna that
the two limits are combined --- at the amplitude level.  In this section,
we will construct the one-loop analog of the antenna function and the
antenna factorization formula.

\def\Ant{\mathop{\rm Ant}\nolimits}
The antenna function presented in ref.~[\use\Antenna] describes the factorization
of an amplitude when a color-adjacent trio of legs $a$, $1$, and $b$
degenerates into two independent, hard momenta $\ah$ and $\bh$, without
$a$ or $b$ becoming soft.  (This is the limit where 
$\Delta(a,1,b)/s_{ab}^3 \rightarrow 0$, $\Delta$ being the
symmetric Gram determinant of the $(a,1,b)$ system.)  At tree level,
the antenna function is given by the following formula,
$$\eqalign{
\Ant(\ah,\bh\longleftarrow a,1,b) &= 
{1\over s_{a1}}\pol_\bh\tcdot\pol_b \,V_3(a,1,\ah)
+{1\over s_{b1}}\pol_\ah\tcdot \pol_a\,V_3(1,b,\bh)\,,\cr
}\eqn\TreeAntFactor$$
where $V_3$ is the color-ordered gluon three-point vertex.
In addition, we must specify a pair of {\it reconstruction\/}
functions $k_{\ah,\bh} = f_{\ah,\bh}(k_a,k_1,k_b)$
such that $(k_\ah,k_\bh)\rightarrow -(k_a+k_1,k_b)$ sufficiently quickly
when $k_1\parallel k_a$,
and likewise $(k_\ah,k_\bh)\rightarrow -(k_a,k_b+k_1)$ 
sufficiently quickly when $k_1\parallel k_b$.  These
functions are not unique; different choices will lead to
functional forms for the antenna function that differ by
terms non-singular in any of the singular limits.  We can
also express the tree-level antenna function in terms of the
tree-level splitting amplitude as follows,
$$\eqalign{
\Ant^\tree(\ah,\bh\longleftarrow a,1,b) &= 
-\pol_\bh\tcdot\pol_b \,\Split^\tree(\ah\rightarrow a\,1; 1-{s_{1b}\over K^2})
-\pol_\ah\tcdot \pol_a\,\Split^\tree(\bh\rightarrow 1\,b;{s_{a1}\over K^2})\,,\cr
}\eqn\TreeAntFactorB$$
where $K^2 = s_{\ah\bh} = (k_a+k_1+k_b)^2$.

By analogy, we may now define the following function,
$$
L(\ah,\bh\longleftarrow a,1,b) = 
-\pol_\bh\tcdot\pol_b \,\Cone(\ah\rightarrow a\,1; 1-{s_{1b}\over K^2})
-\pol_\ah\tcdot \pol_a\,\,\Cone(\bh\rightarrow 1\,b; {s_{a1}\over K^2})\,.
\anoneqn$$

What are the limits of this function in the three singular regions,
$k_a\parallel k_1$, $k_b\parallel k_1$, and $k_1\rightarrow 0$?

In the first region, $\Cone(\bh\rightarrow 1\,b;z)$ has at worst
a logarithmic singularity as $s_{a1}\rightarrow 0$.  
Furthermore, $k_b$ can be thought of as a reference momentum for the
purposes of defining $z$,
$$
z = {s_{ab}\over s_{1b}+s_{ab}} \simeq {s_{ab}\over s_{a1}+s_{1b}+s_{ab}}\,.
\anoneqn$$
The $L$ function thus reduces to the ordinary one-loop splitting amplitude
in this region.  Similarly, it reduces to the splitting amplitude in the
region $k_b\parallel k_1$.

In the soft region, $k_1\rightarrow 0$,
$$\eqalign{
\Ctree(\ah\rightarrow a\,1; z) &\rightarrow {\sqrt2\over s_{a1}} 
        k_a\cdot\pol_1\,\pol_a\cdot\pol_{\ah}\,,\cr
\Ctree(\bh\rightarrow 1\,b; z) &\rightarrow -{\sqrt2\over s_{1b}} 
        k_b\cdot\pol_1\,\pol_b\cdot\pol_{\bh}\,,\cr
}\anoneqn$$
while the second term in eqn.~(\use\GeneralFormA) is finite,
and thus may be dropped.  In taking the limit of $L$, we may set the
relevant $z$ to either zero or one as appropriate, except in expressions
of the form $z^\e$.
The leading behavior of $L$ in the soft limit is thus given by
$$\eqalign{
&-{1\over2} \left\{\L{\mu^2\over -s_{a1}}\R^\e
\,\LB  {s_{1b}\over K^2} f_1\L{s_{1b}\over K^2}\R
+f_1\L 1-{s_{1b}\over K^2}\R -2 f_2\RB\,
{\sqrt2\over s_{a1}} k_a\cdot\pol_1\,\pol_a\cdot\pol_{\ah}\,\pol_b\cdot\pol_{\bh}
\RP\cr &\hskip 10mm \LP
-\L{\mu^2\over -s_{1b}}\R^\e
\,\LB  {s_{1a}\over K^2} f_1\L{s_{1a}\over K^2}\R
+f_1\L 1-{s_{1a}\over K^2}\R -2 f_2\RB\,
{\sqrt2\over s_{1b}} k_b\cdot\pol_1\,\pol_a\cdot\pol_{\ah}\,\pol_b\cdot\pol_{\bh}
\right\} \,;
}\anoneqn$$
using 
$$
f_1(z) \buildrel{z\rightarrow 1}\over\sim
{2 \over\e^2}\cg \LB -{\Gamma(1-\e)\Gamma(1+\e)}
                            \, (1-z)^\e
                        -1
                        +(1-z)^\e\,\Gamma(1+\e)\Gamma(1-\e)\RB
 = 2 f_2\,,
\anoneqn$$
and
$$
z f_1(z) \buildrel{z\rightarrow 0}\over\sim
 -{2\over\e^2} \cg \Gamma(1+\e)\Gamma(1-\e) z^{-\e}\,,
\anoneqn$$
we obtain ($K^2 = s_{ab}$ in the limit)
$$\eqalign{
&{\cg \over\e^2}\L{\mu^2 (-s_{ab})\over (-s_{a1})(- s_{1b})}\R^\e
\Gamma(1+\e)\Gamma(1-\e)
\,\pol_a\cdot\pol_{\ah}\,\pol_b\cdot\pol_{\bh}
\Bigl( {\sqrt2\over s_{a1}} k_a\cdot\pol_1
- {\sqrt2\over s_{1b}} k_b\cdot\pol_1\Bigr)
\cr &= 
-{\cg \over\e^2} \L{\mu^2 (-s_{ab})\over (-s_{a1})(- s_{1b})}\R^\e
\Gamma(1+\e)\Gamma(1-\e)
\,\pol_a\cdot\pol_{\ah}\,\pol_b\cdot\pol_{\bh}
\,\Soft^\tree(a,1,b)\,.
\cr &= 
-{\cg \over\e^2} \L{\mu^2 (-s_{ab})\over (-s_{a1})(- s_{1b})}\R^\e
{\pi\e\over\sin\pi\e}
\,\pol_a\cdot\pol_{\ah}\,\pol_b\cdot\pol_{\bh}
\,\Soft^\tree(a,1,b)\,.
}\anoneqn$$
The latter form agrees with the result given in ref.~[\use\BDS].

\def\Ant{\mathop{\rm Ant}\nolimits}
The function we have defined is thus the appropriate generalization
of the antenna function to one loop,
$$\Ant^\oneloop(\ah,\bh\longleftarrow a,1,b) = 
L(\ah,\bh\longleftarrow a,1,b)\,.
\anoneqn$$
  With it in hand,
we may summarize the singular behavior of a leading-color one-loop 
color-ordered amplitude
whenever the trio of momenta ($a,1,b$) become degenerate without
either $a$ or $b$ becoming soft as follows,
$$\eqalign{
 A_n^{\oneloop}&(1,\ldots,a^\la,b^\lb,\ldots,n) 
\;{\buildrel \Delta(a,1,b)/s_{ab}^3 \rightarrow 0
    \over{\relbar\mskip-1mu\joinrel\longrightarrow}}\cr
&\sum_{\phpol\ (\ah,\bh)}  \biggl(
  \Ant^\tree(\ah,\bh\longleftarrow a,1,b)\,
\, A_{n-1}^\oneloop(1,\ldots,-\ah(a,1,b),-\bh(b,1,a),\ldots,n)
\cr &\hskip 20mm  
  +\Ant^\oneloop(\ah,\bh\longleftarrow a,1,b)\,
\, A_{n-1}^\tree(1,\ldots,-\ah(a,1,b),-\bh(b,1,a),\ldots,n)\,\biggr) \;.
}\eqn\OneLoopAntennaFactorization$$

As is the case at tree-level, this function contains
terms which 
are not singular in any of the soft or collinear limits.
Such terms lead to finite contributions when integrated
over `singular' regions of phase space.  The inclusion
or exclusion of such terms is arbitrary; in a physical observable
 they will cancel between `virtual+singular' contributions
and `real-emission' contributions.

\section{Conclusions}
\vskip 10pt

The cut-based method has been used 
extensively in recent years for computations of various gauge-theory
amplitudes.  We have shown here that it can also be used for
simple, direct computations of the universal functions governing
the collinear behavior of loop amplitudes, to all orders in the
dimensional regulator $\e$.  We have also shown how to combine
the different singular limits of real emission in one-loop
amplitudes --- soft and collinear --- into
a single {\it antenna\/} function.  A related function introduced
by Catani and Seymour~[\use\CS] for tree-level amplitudes, 
the dipole factorizing function, has already proven useful in 
a formalism for handling and cancelling infrared divergences
in next-to-leading order calculations.  The antenna function
presented here should be similarly useful in next-to-next-to-leading
order computations of jet processes in perturbative QCD.

We thank Z. Bern, V. Del Duca, W. B. Kilgore, and C. R. Schmidt
for an advance copy of a preprint~[\use\BDKS] on another method for performing
the calculations discussed here.

\appendix{Integrals}
\tagappendix\IntegralsAppendix
\vskip 10pt

In addition to the linear tensor integral, $\AI_3^\mu$, given in 
section~\use\ExampleSection, for the calculations in section~\use\GeneralSection,
we also need the following integrals,
$$\eqalign{
\AI_4^\mu(s_{ab},z) &= -i\int {d^{4-2\e} p\over (2\pi)^{4-2\e}}\;
   {p^\mu\over p^2 (p-k_a-k_b)^2 \,p\cdot q}\cr
&= {1\over \L -s_{ab}\R^{\e}\,q\cdot (a+b)}\,\LB
       -{\e\over (1-2\e)} f_2\,(k_a+k_b)^\mu 
       +{1\over (1-2\e)} {s_{ab}\over 2 q\cdot (a+b)} f_2 q^\mu\RB\,,\cr
\AI_5^{\mu\nu}(s_{ab},z) &= -i\int {d^{4-2\e} p\over (2\pi)^{4-2\e}}\;
   {p^\mu p^\nu\over p^2 (p-k_a)^2 (p-k_a-k_b)^2 \,p\cdot q}\cr
&= {1\over \L -s_{ab}\R^{1+\e}\,q\cdot (a+b)}\,\LB
       f_{5aa}(z) k_a^\mu k_a^\nu 
       + {1\over 2}f_{5ab}(z) \L k_a^\mu k_b^\nu + k_b^\mu k_a^\nu\R
       + f_{5bb}(z) k_b^\mu k_b^\nu
\RP\cr&\hskip 10mm\LP
  +{s_{ab}\over 4 q\cdot (a+b)} f_{5aq}(z) \L k_a^\mu q^\nu+q^\mu k_a^\nu\R
  +{s_{ab}\over 4 q\cdot (a+b)} f_{5bq}(z) \L k_b^\mu q^\nu+q^\mu k_b^\nu\R
\RP\cr&\hskip 10mm\LP
        + {s_{ab}^2\over 4 \L q\cdot (a+b)\R^2} f_{5qq}(z) q^\mu q^\nu
        + s_{ab} f_{5g}(z) g^{\mu\nu}
        \RB\cr
}\anoneqn$$
where
$$\eqalign{
f_{5aa}(z) &= {(1 - \e)\over 2 (1 - 2 \e)} f_1(z)
  \,,\cr
f_{5ab}(z) &= 
    -{(1 - \e) z\over{(1 - 2 \e) (1 - z)}}  f_1(z) + 
     {2 (1 - \e) \over(1 - 2 \e) (1 - z)} f_2(z)
  \,,\cr
f_{5bb}(z) &= {(1 - \e) z^2 \over2 (1 - 2 \e) (1 - z)^2} f_1(z) 
       - {(\e + z - 2 \e z) \over(1 - 2 \e) (1 - z)^2} f_2(z)
  \,,\cr
f_{5aq}(z) &= -{\e \over(1 - 2 \e) (1 - z)} f_1(z) 
     + {2 \e \over(1 - 2 \e) (1 - z)} f_2(z)
  \,,\cr
f_{5bq}(z) &= -{(1 - \e) z \over(1 - 2 \e) (1 - z)^2} f_1(z)
    + {2 (1 - \e z) \over(1 - 2 \e) (1 - z)^2} f_2(z)
  \,,\cr
f_{5qq}(z) &= {(1 - \e) \over2 (1 - 2 \e) (1 - z)^2} f_1(z)
     - {(2 - \e - z) \over(1 - 2 \e) (1 - z)^2} f_2(z) 
  \,,\cr
f_{5g}(z) &= {z \over4 (1 - 2 \e) (1 - z)}f_1(z) - {1\over2 (1 - 2 \e) (1 - z)} f_2(z)
  \,.\cr
}\anoneqn$$

For completeness, we also list the ordinary tensor bubble and triangle
integrals of which we make use.  These are of course independent of 
the collinear momentum fraction $z$, and may therefore be expressed using
$f_2$,
$$\eqalign{
I_2 &= -i\int {d^{4-2\e} p\over (2\pi)^{4-2\e}} \; {1\over p^2 (p-k_a-k_b)^2}
\cr    &= -{\e\over (1-2\e)} {1\over (-s_{ab})^\e}\, f_2\,,
\cr
I_{2a}^\mu &= -i\int {d^{4-2\e} p\over (2\pi)^{4-2\e}} \; {p^\mu\over p^2 (p-k_a-k_b)^2}
\cr    &= -{\e\over 2 (1-2\e)} {1\over (-s_{ab})^\e}\, f_2\, (k_a+k_b)^\mu\,,
\cr
I_{2b}^{\mu\nu} &= 
   -i\int {d^{4-2\e} p\over (2\pi)^{4-2\e}} \; {p^\mu p^\nu\over p^2 (p-k_a-k_b)^2}
\cr &= {\e\over2 (3-2\e)(1-2\e)} {1\over (-s_{ab})^\e}\,f_2\,
  \LB {1\over2} s_{ab} g^{\mu\nu}
      - {(2-\e)} (k_a+k_b)^\mu (k_a+k_b)^\nu\RB\,,\cr
I_3 &= 
   -i\int {d^{4-2\e} p\over (2\pi)^{4-2\e}} \; {1\over p^2 (p-k_a)^2 (p-k_a-k_b)^2}
\cr &= {1\over \L -s_{ab}\R^{1+\e}}\,f_2\,,\cr
I_{3a}^\mu
&= -i\int {d^{4-2\e} p\over (2\pi)^{4-2\e}}\; 
       {p^\mu\over p^2 (p-k_a)^2 (p-k_a-k_b)^2}\cr
&= {(1-\e)\over (1-2\e)}
  \,(-s_{ab})^{-1-\e}\, f_2 k_a^\mu
   -{\e\over (1-2\e)}
  \,(-s_{ab})^{-1-\e}\, f_2 k_b^\mu\,,\cr
I_{3b}^{\mu\nu} &= 
   -i\int {d^{4-2\e} p\over (2\pi)^{4-2\e}} \; 
            {p^\mu p^\nu\over p^2 (p-k_a)^2 (p-k_a-k_b)^2}\cr
&= {1\over \L -s_{ab}\R^{1+\e}}\,{\e \over 4 (1-\e)(1-2\e)}s_{ab}\,f_2\,g^{\mu\nu}
\cr &\hskip5mm +{(2-\e)\over2\,(1-2\e)}
  \,{1\over \L -s_{ab}\R^{1+\e}} \,f_2\,
       \L k_a^\mu k_a^\nu - {\e\over 1-\e}(k_a^\mu k_b^\nu + k_b^\mu k_a^\nu)
          -{\e\over 2-\e} k_b^\mu k_b^\nu\R\,,\cr
I_{3c}^{\mu\nu\rho} &= 
   -i\int {d^{4-2\e} p\over (2\pi)^{4-2\e}} \; 
            {p^\mu p^\nu p^\rho\over p^2 (p-k_a)^2 (p-k_a-k_b)^2}\cr
&= {1\over \L -s_{ab}\R^{1+\e}}\,{\e (2-\e)\over 4 (1-\e)(1-2\e)(3-2\e)}s_{ab}\,f_2\,
      \L g^{\mu\nu} k_a^{\rho} +g^{\mu\rho} k_a^{\nu}+g^{\nu\rho} k_a^{\mu}\R
\cr&\hskip 5mm
+{1\over \L -s_{ab}\R^{1+\e}}\,{\e \over 4 (1-2\e)(3-2\e)}s_{ab}\,f_2\,
      \L g^{\mu\nu} k_b^{\rho} +g^{\mu\rho} k_b^{\nu}+g^{\nu\rho} k_b^{\mu}\R
\cr &\hskip5mm 
   +{(2-\e)(3-\e)\over 2(1-2\e)(3-2\e)}
  \,{1\over \L -s_{ab}\R^{1+\e}} \,f_2\,
\cr &\hskip10mm\times
         \Bigl( k_a^\mu k_a^\nu k_a^\rho
           -{\e\over 1-\e} (k_a^\mu k_a^\nu k_b^\rho +k_a^\mu k_b^\nu k_a^\rho
                              +k_b^\mu k_a^\nu k_a^\rho)
\cr&\hskip 15mm
           -{\e\over 2-\e} (k_a^\mu k_b^\nu k_b^\rho +k_b^\mu k_a^\nu k_b^\rho
                              +k_b^\mu k_b^\nu k_a^\rho)
           -{\e\over 3-\e} k_b^\mu k_b^\nu k_b^\rho\Bigr)\,.\cr
}\anoneqn$$

\listrefs
\bye